\documentclass[11pt]{article}
\pdfoutput=1
\usepackage[inline]{enumitem}
\usepackage{amsfonts}
\usepackage{amsmath}
\usepackage{amssymb}
\usepackage{array}
\usepackage{caption}
\usepackage{caption}
\usepackage{centernot}
\usepackage{color}
\usepackage{comment}
\usepackage{dcolumn}
\usepackage{datetime}
\usepackage{diagbox}
\usepackage{eurosym}
\usepackage{footmisc}
\usepackage{geometry}
\usepackage{graphicx}
\usepackage{makecell}
\usepackage{multirow}
\usepackage{multicol}
\usepackage{natbib}
\usepackage{pdflscape}
\usepackage{rotating}
\usepackage{sectsty}
\usepackage{setspace}
\usepackage{standalone}
\usepackage{subfigure}
\usepackage{tikz}
\usepackage{tikz-qtree}
\tikzset{
  hollow node/.style={circle,draw,inner sep=2},
  empty  node/.style={rectangle,draw,fill=white,color=white},
}
\usepackage{titling}
\usepackage{ulem}
\usepackage[hidelinks]{hyperref}
\hypersetup{unicode = true}
\newcommand{\T}{\rule{0pt}{2.6ex}}            % Top strut
\newcommand{\B}{\rule[-1.2ex]{0pt}{0pt}}      % Bottom strut

\newcommand{\subtitle}[1]{\posttitle{\par\end{center}\begin{center}\large#1\end{center}\vskip0.5em}}
\newcommand{\refp}[1]{(\ref{#1})}

\begin{document}

\begin{titlepage}
\title{
  Electoral Crime Under Democracy: Information Effects from Judicial Decisions in Brazil\thanks{
    I am grateful to the insightful feedback from Brigitte Seim, Ashu Handa, Scott Desposato, Tricia Sullivan, Doug MacKay, Patricio Dominguez, Guilherme Lambais, Lara Mesquita, George Avelino, Ciro Biderman, Henrik Sigstad, Juan Carlos Salgado, Julio Trecenti, and seminar participants at CEPESP-FGV, the Paraná State Bar Association, the 2019 Global Conference on Transparency Research, the Harvard Kennedy School, the Institute for Humane Studies, and the 2019 APPAM Fall Conference. I acknowledge financial support from the Institute for Humane Studies (IHS) and the UNC-Chapel Hill Department of Public Policy. All remaining errors are my own.
  }
}
\author{
  Andre Assumpcao\thanks{
    Department of Public Policy, The University of North Carolina at Chapel Hill; \href{mailto:aassumpcao@unc.edu}{\textcolor{blue}{aassumpcao@unc.edu}}
  } %\textsuperscript{,}\thanks{
    %Center for International Development, Harvard Kennedy School.
  %}
}
\date{\monthname[\month], \the\year}
\maketitle

\begin{abstract} \onehalfspacing
This paper examines voters' responses to the disclosure of electoral crime information in large democracies. I focus on Brazil, where the electoral court makes candidates' criminal records public before every election. Using a sample of local candidates running for office between 2004 and 2016, I find that a conviction for an electoral crime reduces candidates' probability of election and vote share by 10.3 and 12.9 percentage points (p.p.), respectively. These results are not explained by (potential) changes in judge, voter, or candidate behavior over the electoral process. I additionally perform machine classification of court documents to estimate heterogeneous punishment for severe and trivial crimes. I document a larger electoral penalty (6.5 p.p.) if candidates are convicted for severe crimes. These results supplement the information shortcut literature by examining how judicial information influences voters' decisions and showing that voters react more strongly to more credible sources of information.\\
\vspace{0in} 

\vspace{0in}
\bigskip
\end{abstract}

\setcounter{page}{0}

\thispagestyle{empty}

\end{titlepage}

\pagebreak

\section{Introduction} \label{sec:introduction_paper1}

In democratic regimes, office-seeking politicians employ various tactics to get elected. They might promise voters the provision of more local public goods, such as schools, hospitals, or roads; they promote their candidacies by running ads on TV and, more recently, on social media; they might even meet with their constituents and ask for their vote based on their personal connection. While these tactics are different, sometimes complementary ways to win an election, they all characterize legal strategies, in which individuals follow the law when running for office. Governments allow such electoral practices because they make electoral systems more competitive, and increase access to elected office. Nevertheless, two important questions remain as to how illegal practices affect elections in democratic regimes and how voters react when these practices are made public. In this paper, I answer these questions by analyzing municipal elections and court decisions in Brazil.

Scholars have not ignored the use of illegal strategies in elections. \citet{LehoucqElectoralFraudCauses2003} offers a comprehensive account of the relationship between electoral fraud and election outcomes, discussing the various forms of crime, such as procedural rule-breaking, illegal campaigning, violence, and vote buying. In a more recent study, \citet{Gans-MorseVarietiesClientelismMachine2013a} design a theoretical framework encompassing four types of clientelism practices (vote, turnout, and abstention buying, and double persuasion) and their adoption under five different institutional designs. They argue that the choice of illegal action is conditional on the design of the electoral system. For instance, in an environment of increased political polarization, we should expect to see more of turnout buying but less of vote buying. However, the majority of these studies looking into electoral crimes have limited scope: first, they are primarily concerned with coercive threats that prevent free and fair elections \citep[as suggested by][]{MaresBuyingExpropriatingStealing2016}; second, they focus heavily on non or partially democratic regimes, evidenced by the vast literature on electoral authoritarianism in \citet{AsunkaElectoralFraudViolence2017a,GandhiElectionsAuthoritarianism2009,IchinoDeterringDisplacingElectoral2012,LevitskyRiseCompetitiveAuthoritarianism2002,SchedlerElectoralAuthoritarianism2015}.

The first contribution of this paper is examining electoral crimes through an information shortcut angle. I posit that rulings issued by electoral courts on claims of electoral crime serve as signals of candidate type that help citizens make voting decisions. In a crowded candidate space, judicial sentences help distinguishing good and bad candidates. The information shortcut literature is well-established \citep[and others]{PopkinReasoningVoterCommunication1991,LupiaElementsReasonCognition2000}, but to my knowledge this is one of the first attempts at using court documents as sources of candidate information. Brazil's recent democratic history and open data environment create a unique context in which the electoral process, including electoral litigation, are very visible during campaign periods, and thus play an important role in voting decisions. This paper also uncovers the effect of electoral crimes that are harder to detect or whose relationship with electoral outcomes is less known or well understood. For instance, politicians might use illegal forms of advertising or slush funds to spend beyond their campaign limits in order to win an election. In democratic regimes, we are more likely to experience these illegal strategies than coercive threats or flat out vote buying, for example.

Besides the electoral fraud and information cues scholarship, the present study also advances the broader literature on the political economy of development. Brazil has a unique institutional design in which the judiciary branch has an entire system of State (TRE) and Federal (TSE) electoral courts resolving electoral disputes. Their mandate is to guarantee free and fair competition for public office, enforcing the Brazilian Electoral Code of 1965 and subsequent legislation, and to prevent that candidates not meeting legal requirements join electoral races. To the extent that electoral courts are successful in rooting out this type of wrongdoing, we should expect more electoral accountability from office-seeking politicians. Candidates would also avoid breaching electoral law to preserve their future career prospects. Understanding if electoral oversight mechanisms as such are effective should provide an important takeaway for countries with similar institutions. Mexico, India, and South Africa are but a few developing countries which also have a dedicated electoral authority similar to Brazil's. These monitoring institutions are an important source of judiciary power beyond their traditional role of settling legal disputes; since every political candidate in Brazil needs a judicial authorization to run for office, electoral courts hold an enormous amount of power in shaping up political representation -- an unusual role played by the judiciary branch in developing countries.

The other contributions of this paper are methodological in nature. First, I use court documents as data. I collect and code judicial rulings on candidacy authorization cases for individuals running for municipal office in Brazil between 2004 and 2016. For a subset of these documents, I implement multiple machine classification algorithms to find the exact allegations against candidates that prevent them from running for office. I split such charges into two categories, severe and trivial rule-breaking, to identify heterogeneous information effects of electoral crime on performance for the 2012 and 2016 elections. This approach aligns with a recent wave of studies using court documents to measure economic and political outcomes in development settings \citep{Sanchez-MartinezDismantlingInstitutionsCourt2017,LambaisJudicialSubversionEvidence2018,Poblete-CazenaveCrimePunishmentPoliticians2019}.

Using these court documents, I recover the causal effect of criminal information on voter responses adopting an instrumental variables (IV) strategy. For a subsample of candidates accused of electoral crimes in Brazil, I observe (endogenous) trial decisions issued before elections and (exogenous) appeals decisions issued after elections. These candidates remained in their race and could be voted for on election day. While trial rulings are endogenous, e.g.,~potentially correlated with other factors determining a candidate's electoral chances, appellate decisions issued \emph{after} election day cannot influence electoral outcomes. Thus, for this subsample of candidates who have untried appeals standing at the time of election, I can identify the causal effect of criminal information on their electoral performance.

The main IV result shows that revealing criminal information, i.e., convictions in court for electoral crimes, reduces the probability of election and a candidate's vote share by 10.3 and 12.9 percentage points, respectively. These estimates are statistically significant at the one percent level and significantly differ from OLS estimates. These results are robust to the inclusion of covariates and fixed-effects \iffalse, coefficient stability tests \citep{OsterUnobservableSelectionCoefficient2019}, \fi and inclusion or exclusion restriction checks. Convicted candidates are also significantly further away from the vote threshold necessary for election in both proportional (city council) and majoritarian (mayor) systems. The significant effect of judicial convictions means voters interpret them as negative signs of candidate type and hand out punishment at the ballot box. This is a novel information shortcut effect, and it supplements party label, interest group, and media effects previously identified in the literature. It also closely aligns with similar studies documenting a negative effect of corruption information on voting decisions in Brazil \citep[and others]{FerrazExposingCorruptPoliticians2008b,FerrazElectoralAccountabilityCorruption2011a,WintersLackingInformationCondoning2013}.

I test multiple alternative explanations for the effect on electoral performance. I show that (potential) changes in judicial sentencing behavior, campaign strategies, and voter engagement cannot explain the hit to electoral performance, leaving punishment for electoral crimes as the only plausible explanation for the negative information effect. I further find that the punishment for severe violations (e.g., candidates or parties using illegal campaign strategies, channeling slush funds for campaign ads, having previous outstanding judicial convictions preventing them from running for office) is harsher than for trivial violations, indicating that voters distinguish illegal behavior and fit punishment to the type of crime. In \citet{Weitz-ShapiroCanCitizensDiscern2017}, voters state their dislike for corruption more strongly when the criminal information comes from more credible sources; here, I show that voters act on criminal information originating from the most respected branch of government in Brazil. Thus, I document that voters' follow through on their stated preferences, supplementing the evidence in \citet{Weitz-ShapiroCanCitizensDiscern2017}. When candidates are cleared of severe accusations, however, they experience an improvement in their electoral performance, likely to due to the ability of these electoral strategies to sway voters in their favor. This result indicates a positive payoff of engaging in illegal strategies as long as candidates are cleared by the courts.

In the remainder of this paper, I discuss the theoretical framework for the effect of criminal information in section \ref{sec:theory_paper1}, and explain the institutional background in Brazil allowing for causal identification in section \ref{sec:background_paper1}. I present the data in section \ref{sec:data_paper1}. Section \ref{sec:methods_paper1} discusses the empirical strategy of this paper, and section \ref{sec:results_paper1} presents its main results. Section \ref{sec:exclusionrestriction_paper1} explores the exclusion restriction tests. In section \ref{sec:alternatives_paper1}, I investigate alternative explanations for the negative effect on performance. Section \ref{sec:hte_paper1} discusses the heterogeneous punishment effects. Section \ref{sec:conclusion_paper1} concludes and suggests further avenues of research.

\section{Court Signals of Electoral Crimes} \label{sec:theory_paper1}

Assume three representative agents are interacting in a democratic election: the candidate ($C$), the voter ($V$), and the judge ($J$). They each maximize their utility function $f(X_c, \, \varepsilon_c)$, which summarizes their electoral response to a matrix of observed and unobserved candidate characteristics ($X_c$ and $\varepsilon_c$). The former could be anything from policy positions, age, ethnicity, marital status, or campaign expenditures. A candidate's political ability, the deals they make with parties, supporters, or sponsors are the latter. The more information voters have, the better they can choose their preferred candidates. \citet{PopkinReasoningVoterCommunication1991} suggested that voters gather available signals to form an opinion about candidates in an election. For instance, $V$ might prefer highly-educated candidates, so $V$ seeks signals about a candidate's educational background. Generally, however, $V$ dislikes candidates who have a criminal record because it signals dishonesty and indicates poor prospective political performance. Thus, I separate criminal information $c_{c}$ from matrix $X_{c}$ in \emph{V}'s utility function and set the first derivative of $f_{V}$ with respect to $c_{c}$ to negative, as follows in equations \refp{eq:1} and \refp{eq:2}: \vspace{-10pt}
\begin{multicols}{2}
  \begin{equation} \label{eq:1}
    U_{V} = f_{V}(X_{c}, c_{c}, \, \varepsilon_{c})
  \end{equation}
  \vfill
  \columnbreak
  \begin{equation} \label{eq:2}
    \frac{\partial U_{V}}{\partial c_{c}} < 0
  \end{equation}
\end{multicols}

In addition to the representative voter's preferences, I am also interested in candidate \emph{C}'s behavior. \emph{C} is looking to adopt strategies that maximize their electoral outcome. They cannot withhold or control specific information, such as age, gender, ethnicity, but can choose amongst campaign expenditure levels (included in $X_{c}$) and electoral strategies that get them closer to winning an election. These strategies could be anything from running ads on TV and social media, holding campaign events, or paying someone off the books to find dirt on their opponents. Clearly, the last strategy is illegal, and candidates want to hide it from the public. Voters find illegal strategies a dishonest move and would likely condemn such practices; candidates, however,  would still want to adopt them because they might shift  support away from opponents and thus compensate the potential loss from disclosure of criminal action. Therefore, \emph{C} adopts a mix of strategies such that their expected electoral payoff remains positive. Illegal strategies enter the candidate's utility function as $b_c$.
\begin{equation} \label{eq:3}
  U_{C} = f_{C}(X_{c}, c_{c}, b_{c}, \, \varepsilon_{c})
\end{equation}

% I assume a uniform distribution of detection risk across electoral districts for the reasons laid out in section \ref{sec:background_paper1}, guaranteeing the independence and quality of electoral judges in Brazil. Note that applications of this simple model to other jurisdictions would likely change this assumption to adjust to features of other judiciary systems.
The focus in this paper is identifying how the disclosure of criminal records and illegal strategies, respectively summarized by $c_{c}$ and $b_{c}$, impact a candidate's chances of election. More specifically, I am looking at the effect of releasing electoral crime information on electoral performance. The information could come from the media, an important source of candidate signals, but it could also come from government authorities, such as the judiciary branch. The reasoning is straightforward. Assuming an independent, high-quality judiciary, judge $J$ hands out sentences based on case evidence, either convicting or acquitting candidates, regardless of individual characteristics. Since voters dislike crimes, and judges make criminal information available to voters by ruling on candidate cases, one can reasonably expect voters to interpret judicial decisions as signals of candidate type and respond accordingly at the ballot. A conviction is a negative signal; an acquittal is a positive signal.

This mechanism would predict a negative first derivative $\partial U_{C} / \partial c_{c} < 0$ for $c_{c}$ in equation \refp{eq:3}: convictions on record hurt a candidate's chances to hold office. Some of this effect, however, could be offset by the boost in votes that would come from illegal strategies $b_{c}$. Suppose a candidate prints and distributes negative material on their opponents. The information in the advertisement is false, and such ad is not allowed in the jurisdiction where this office race is taking place. Though judges could eventually ban such material, once the information is out, it might hurt targeted opponents beyond reparation. In such hypothetical scenario, the strategy was illegal but benefited the candidate running the ad. In this case, $ \partial U_{C} / \partial b_{c} > 0$. I want to test both effects on electoral performance.

% \subsection{Application to the Brazilian Context} \label{subsec:brazil_paper1}

\section{Electoral Court System in Brazil} \label{sec:background_paper1}

The majority of the literature in electoral crimes is concerned with more flagrant violations of elections, such as fraud or vote buying \citep{LehoucqElectoralFraudCauses2003}. This paper is an important contribution to the scholarship by looking at other, more common, and more nuanced violations to electoral rules that are particular to large democracies, such as Brazil. To the extent that voters interpret the judicial information as a signal of candidate type, we should expect a change in voter behavior as a result of the disclosure of information.

Brazil is a particularly interesting research setting because of the structure of its dedicated electoral court. Federal (TSE) and State (TRE) Electoral Court systems have existed intermittently since 1932 but only became institutionally relevant after the country's return to democracy in 1985. Since then, electoral courts have a fundamental role in guaranteeing free and fair elections. Their mandate is to enforce the Electoral Code of 1965 and subsequent legislation, particularly the Law Establishing Conditions for Ineligibility to Public Office (1990), the Law of Political Parties (1995), the Law of Elections (1997), and the Clean Records Act of 2010. These courts have four primary responsibilities: (i) electoral rule-making; (ii) judicial consultations clarifying and establishing jurisprudence for conflicting electoral norms; (iii) administration of the electoral process, which consists in publishing the electoral calendar, testing voting machines, distributing voting machines to all districts, counting votes and publishing electoral results; and, finally, (iv) conflict resolution on claims of breach of electoral law.

In this project, I am interested in the courts' conflict resolution function and its underlying judicial review process. According to Brazilian Law, every individual running for office, at every level, has to submit proper documentation proving that they meet eligibility requirements for the office to which they are running; for instance, they should be 35 years of age or older to run for president or senator; executive-office holders, if running for any other elected office, must step down from their current post six months before election day. Every electoral cycle, the highest-level electoral court, TSE, establishes a calendar for submission of all these documents, which are reviewed at lower-level courts by electoral judges who issue rulings authorizing every single candidacy in the country. These cases are called \emph{registro de candidatura} (candidacy registration).

Importantly, these cases follow the usual judicial review process in Brazil. Local electoral judges rule on every candidacy in the country and issue trial decisions either authorizing or dismissing candidacies. Judges check whether a candidate's party has met all electoral requirements, whether candidates have met all criteria of the office to which they are running, and other legal provisions as established by electoral law. Candidates, opponents, or the Office of Electoral Prosecution (MPE) can then appeal trial decisions. An appellate panel of electoral judges at the state level, instead of the district level, reviews the appeal and either affirms or reverses the trial decision. Since appeals can be filed by opponents or the MPE, candidacy registration cases are unique in that they allow us to observe favorable and unfavorable decisions at both stages. Using the timing according to which these decisions are issued, I can create exogenous variation in the disclosure of electoral crimes and analyze their impact on elections.

\subsection{Timing of Judicial Review as Exogenous Shock} \label{subsec:causal}

An example helps detailing how the timing of decisions shocks the disclosure of electoral crimes. The most recent municipal election took place on Oct 2, 2016. The deadline for submitting all candidacy documents was Aug 15, 2016. Between Aug 15 and Sep 12, trial courts reviewed and authorized every candidacy for mayor or city council. The review process started at the electoral district in which the candidate is running for office, and their trial ruling came out of the designated electoral judge for that district. These judges are part of the state court system and, when appointed to the electoral bench, are on leave from their original tenured positions at the state system.\footnote{In Brazil, judges are appointed to the bench in state and federal courts when they pass nationally-competitive entrance examinations. They are automatically tenured after a two-year trial period; therefore, their entire career is independent of electoral politics.} They serve on two-year mandates, with one reappointment allowed, such that they never oversee the same district for more than one electoral cycle. When either a candidate or someone else, such as opponents or the MPE, filed an appeal to the trial ruling, the case was presented before a panel of three judges at the state electoral court TRE. There are seven appellate court justices in each state's TRE, serving up to four-year mandates, and they are immune to local politics. In any state, six of these judges are voted in by their fellow tenured judges at the state and federal court systems and the last member is appointed by the President of Brazil. If plaintiffs or defendants were unhappy with the appellate court decision, they could have appealed their case before the federal court TSE, which serves as the third and final instance of judicial review for mayor and city council candidates.

The Sep 12\footnote{The exact day varies marginally every cycle. In 2018, for instance, the deadline for candidacy submission was Aug 15, last day for loading candidate information was Sep 17, and election day was Oct 7.} deadline is the main institutional element supporting causal inference, because it allows the observation of electoral performance for politicians who violate electoral rules. It is the last day for entering candidate information onto electronic voting (EV) machines distributed at every single polling station in the country.\footnote{\citet{FujiwaraVotingTechnologyPolitical2015} describes this technology in detail.} All candidates who have untried appeals by this date will have their information loaded, and thus can be voted for, in the EV machines on election day. Because of this feature, I can observe the electoral performance of candidates who eventually are convicted of electoral crimes and compare to candidates who are eventually cleared of these charges. If candidates saw a final ruling before Sep 12, or if they have decided not to appeal their trial sentence, I cannot observe their performance because TSE will not load their information in the EV machines.\footnote{There is no early voting in Brazil. Voters cast their ballot on a single day using the EV machines.} As such, these candidates are not included in the analysis here.

Exogenous variation in convictions for electoral crimes comes from the timing according to which higher-level courts issue appeals decisions. Often, the high number of candidates running for municipal office, the judicial backlog, or the conditions of a particular electoral race make it difficult for electoral courts to hand out final decisions by Sep 12. Moreover, since candidates with standing appeals will have their information loaded onto EV machines regardless, there is no strong incentive for courts to issue decisions between then and Oct 2. In the lead-up to election day, judges and court officials are working around the clock making sure that 540,000+ working EV machines reach 450,000+ voting stations across the whole territory of Brazil; judges are ruling on smaller electoral cases that might or might not be appended to candidacy cases; court officials are meeting with political parties and discussing the electoral situation, so on and so forth. It is not uncommon, therefore, to see final decisions come out only after election day has passed, especially in municipal elections. When appeals are not ruled in time for elections, the TSE loads candidate information (picture, name, voting number) on the EV machine but their votes are computed \emph{sub judice} -- their vote count will be considered valid only when the TRE or TSE publish their final ruling. Effectively, thus, the decisions at the appeals stage cannot affect electoral outcomes, since they are issued \emph{after} election day has passed, but they bear a strong relationship to the sentence handed out by the trial judge in each electoral district. Decisions at trial are mostly endogenous with respect to local electoral conditions, since unobservable characteristics of each race might determined claims brought up at trial (office competitiveness, visibility of electoral race, media scrutiny, and others), but I can use appeals as instruments to isolate only the exogenous part of trial decisions.

The primary limitation of this study is that I can only recover causal effects of criminal information under restrictive conditions pertaining to municipal elections in Brazil. Local politicians have no control over the careers of electoral judges nor can they sanction judges. At any other electoral race, however, Congress, Senate, and State Governor candidates are much more powerful and can influence how electoral judges rule their cases. For instance, senators are much more influential than city councilors and have a direct channel of communication with the President of Brazil, who is responsible for appointing one judge per TRE. The second limitation is that several candidates do not appeal their trial ruling and as such do not appear on the EV machine on election day. Thus, I cannot observe electoral performance for every candidate who has had their type revealed in the form of convictions for electoral crimes -- just for the subgroup that filed an appeal or has had a third-party appeal their candidacy. It is likely that the latter candidates are heterogeneous in many dimensions when compared to candidates who have not appealed trial decisions, such as their political experience, or their drive to hold elected office. Excluded from this analysis, these candidates should be the object of future projects measuring the effect of judicial (criminal) information on electoral performance in developing countries, and this paper inaugurates such literature.

\subsection{Applying the Model to the Brazilian Context} \label{subsec:brazil_paper1}

Brazil also has a unique open data environment, including court documents, allowing for the investigation of criminal information effects on voter behavior. Information cues are critical factors in democracy because they help voters overcome the rational ignorance problem \citep{NicholsonInformationCuesRational2019}; surprisingly, however, legal sources have not been explored as a factor informing voting decisions. There is research documenting a significant influence on voter behavior from party labels \citep{SamuelsPowerPartisanshipBrazil2014}, candidates' gender \citep{AguilarChoiceSetsGender2015}, their performance \citep{FerrazExposingCorruptPoliticians2008b,WintersLackingInformationCondoning2013,Weitz-ShapiroDiscerningCorruptionCredible2014,WintersWhoChargeHere2016,Weitz-ShapiroCanCitizensDiscern2017}, and even ethnicity when voters have to choose amongst many candidates in large ballots \citep{AguilarBallotStructureCandidate2015}.\footnote{Whether these cues come from identity voting or underlying policy preferences is beyond the scope of this paper. I am only concerned with the extent to which cues shape voting decisions.} Brazil is a new democracy in which the electoral process is particularly popular, with extensive candidate and the electoral process coverage. For better or worse, candidates' litigation history, personal issues, former performance in office, and business arrangements are under the media's scrutiny during campaign months. Therefore, I can use court decisions as cues on candidate type, i.e., whether they are honest or not, and contribute a new information effect to the literature.

More formally, by coding candidacy registration cases, I recover both $c_{c}$ and $b_{c}$ in equation \refp{eq:3}: $U_{C} = f_{C}(X_{c}, c_{c}, b_{c}, \, \varepsilon_{c})$. Sentence outcome, authorizing or dismissing a candidacy, makes up $c_{c}$; legal reasons justifying the sentence make up $b_{c}$. In this study, there are two reasons why judges prevent candidates from running for office: (1) \emph{trivial} rule-breaking, which are cases in which candidates are in breach of electoral law for minor reasons. For instance, they could have forgotten to include a copy of their ID card in their application, or they could have missed a deadline in the application process. In either case, their candidacy is deemed incomplete, and they are not allowed to run for office; or (2) \emph{severe} rule-breaking, which are more egregious cases in which either parties or candidates are in breach of more substantial elements of electoral law. Parties might not have kept, or presented, all financial records from previous elections, candidates might have an outstanding conviction for previous crimes, or they might have been convicted for running illegal campaign strategies against their opponents. These cases are much more likely to be connected to campaign, office, or government crimes disliked by voters.

Another benefit of using candidacy registration cases to identify $c_{c}$ and $b_{c}$ is their standard penalty. Judges disqualify candidacies when they do not meet all requirements, whether the violations are severe or trivial; also, there is no jail time nor immediate financial penalties for candidates and parties, making the legal punishment for candidacy convictions uniform. In general, electoral cases take less time to conclude (17 months on average) than other cases in the Brazilian judicial system (26 months on average) \citep{CNJJusticaEmNumeros2018}. Though standard sentences and penalties might not be ideal from a policymaking perspective, they create a subset of legal cases less susceptible to external influence and relatively stable in terms of the application of legal statutes and convictions.\footnote{These cases, however, are often appended to other cases at the electoral court system and can create financial liabilities for candidates and their parties. The analysis of these other cases is beyond the scope of this study as they do not meet the criteria for causal identification developed here. There is also growing interest for electoral court reform in Brazil. Some experts criticize the fact that electoral justices do not have fixed appointments and thus do not specialize in electoral crimes; others say that parties and candidates strategically avoid harsher punishments by requesting other court systems to move charges to electoral courts, knowing that their punishments are constrained to the electoral arena.}

Finally, these cases allow for the testing of heterogeneous information effects by conviction type. If voters are sophisticated, not only they punish candidates with unfavorable trial rulings (\emph{the conviction, or criminal information, effect}), but they also differentiate the punishment conditional on the crime (\emph{the crime type effect}). One can reasonably expect that candidates charged with more severe crimes, such as illegal campaign spending, or convicted for previous crimes, signal a more systematic criminal behavior and should be punished more harshly than candidates missing deadlines or lacking hard copies of certain documents. Though judicial punishment is the same, the electoral punishment could still reflect the relatively more severe violations. There is substantial evidence in the literature against voter sophistication in other information contexts \citep{AvisGovernmentAuditsReduce2018,Banerjeeinformedvotersmake2010,ChongDoesCorruptionInformation2015,deFigueiredoWhenvoterspunish2011,FerrazElectoralAccountabilityCorruption2011a,Weitz-ShapiroCanCitizensDiscern2017,WintersLackingInformationCondoning2013}; this paper explores yet another mechanism of providing information to voters (judicial decisions) and investigates their reactions.

\section{Data} \label{sec:data_paper1}

The primary data source for electoral performance is TSE's repository of electoral data. TSE publishes electoral results, vote counts, candidates' individual characteristics, and their candidacy's situation on election day for all elections since 1994. I focus on the municipal elections after the introduction of the EV machine in 2002. There are 16,791 candidates for mayor or city council in this sample; these candidates appealed, or had third-parties appealing, the trial ruling on their candidacy case. These candidates were displayed in the EV device and could have been voted for on election day. Their candidacy remained pending after elections; only if a favorable appeals ruling came out were the elected candidates allowed to take up office. I create three outcomes from TSE's data to measure electoral performance: (1) \emph{the probability of election}, which is a binary variable taking up value one when the candidate received enough votes for election. For mayor candidates, under majoritarian rule, this means 50 percent plus one of all valid votes. For city council candidates, under proportional rule, this means having received enough votes to rank amongst the most voted candidates within the designated number of vacancies in each municipality; (2) \emph{vote share} as a share of total valid votes; (3) \emph{vote distance to election cutoff}, which is the percentage point distance between a candidate's vote share and the votes necessary for election. Outcomes (1) and (2) are make or break measures of electoral crime: I can use them to estimate whether a candidate whose type has been revealed as bad is predicted to win or lose an election; conditional on having won (or lost) an election, outcome (3) describes the relative safety (or damage) resulting from the release of criminal information.

Next, I scrape court documents containing the allegations against each candidate from the TSE website. I developed software\footnote{For the benefit of research transparency and replication, all programs and analysis scripts are freely available online on \href{https://www.github.com/aassumpcao/tseresearch}{\textcolor{blue}{GitHub}}.} that downloads case files and sentences for all candidates in my sample. Though the information is public, due to data availability limitations at the TSE, 99.3 percent of court documents come from candidates in the 2012 and 2016 municipal elections. I match court documents to candidates using an individual identifier provided by the Electoral Court so that I can recover all documents for each candidacy.

Table \ref{tab:sumstats} reports the summary statistics of the sample.\footnote{I also run covariate balance tests (available upon request) across convictions at trial to check whether the sample of cleared and convicted candidates have different baseline characteristics. Except for campaign expenditures, there is no significant difference at the 5 percent level across the full set of covariates.} The average age is 46.3 years, and the overwhelming majority of candidates is male. Eight percent of them have any political experience, captured by whether they held any other elected office in the past. These candidates have reported, on average, campaign spending of R\$ 49,924. Using the current exchange rate, this is equivalent to $\sim$\$11,885 per campaign. Fifty-seven percent have seen an unfavorable ruling from the trial judge at their electoral district and 49.7 percent have had an unfavorable ruling after appealing their case to higher courts. Notice that all candidates have seen charges brought against them at trial, otherwise they would not have standing appeals by election day and would not be part of this sample; the conviction variables here capture unfavorable decisions issued by trial judges. If an electoral judge allowed a candidate's run for office, then either the trial or appeals variable becomes zero. Though not reported in table \ref{tab:sumstats}, I also collect information on candidates' marital status and education.\footnote{I also have information on each candidate's party and use it as fixed-effects in the empirical sections.} These are categorical variables, and the most frequent marital status is married (62.6 percent) and education level is high school (30.8 percent). Finally, I report the means for the three outcomes in this analysis. The mean probability of election is 19.4, while the mean candidate's vote share and vote distance to cutoff are 8.2 and $-$8.8 percentage points, respectively.

\section{Empirical Strategy}\label{sec:methods_paper1}

To examine the effect of electoral crimes on electoral performance, the ideal experiment would require a homogeneous sample of candidates differing in only two dimensions: their criminal activity (good politicians have clean records while bad politicians do not) and whether their activity was disclosed by the courts (via convictions or acquittals). Assuming a random distribution of criminal activities and disclosure of records, one could recover causal effects of electoral crime for four groups of candidates:
\begin{enumerate*}[label = (\arabic*)]
 \item bad type disclosed;
 \item bad type undisclosed;
 \item good type disclosed;
 \item good type undisclosed.
\end{enumerate*}
This experiment is unfeasible, because in most cases one cannot observe candidate type neither can one randomize the disclosure of criminal records. The institutional design of candidacy registration cases in Brazil, however, is the best approximation of this experiment. First, all candidates go through criminal background checks as they apply for candidacy in the country. This is a requirement of electoral law, such that we can approximate their type by computing their violations of candidacy requirements. Second, all background checks are public since they are equivalent to sentencing documents and all such documents are public in Brazil.

\begin{figure}[!htbp]
\centering
\caption[figure1]{Instrument Variables Design}\label{fig:ivtree}
% \vspace{-10pt}
\begin{tikzpicture}[
  every level 0 node/.style = {draw=none,anchor=east},
  every level 1 node/.style = {draw,hollow node},
  every level 2 node/.style = {draw,hollow node},
  every level 3 node/.style = {draw,empty node},
  grow                      = right,
  level distance            = .5in,
  sibling distance          = .5in,
  edge from parent path     = {(\tikzparentnode) -- (\tikzchildnode)}
]
\tikzstyle{edge from parent}=[draw,black,very thick]

\begin{scope}[
  yshift = 1.5in,
  align = left,
  edge from parent/.style = {draw=none},
  every level 0 node/.style = {draw=none},
  every level 1 node/.style = {draw=none},
  every level 2 node/.style = {draw=none},
]
\Tree [.{} [.\node[label=left:{Trial Outcome}]{}; [.\node[label=right:{Appeals Outcome}]{}; ]]]
\end{scope}

\Tree [
  .candidacy
  [
    .\node[label=left:{acquittal}]{};
      [.\node[label=right:{conviction (0,1)}]{};]
      [.\node[label=right:{acquittal \hspace{6pt}(0,0)}]{};]
  ][
    .\node[label=left:{conviction}]{};
      [.\node[label=right:{acquittal \hspace{6pt}(1,0)}]{};]
      [.\node[label=right:{conviction (1,1)}]{};]
  ]
]

\end{tikzpicture}
\end{figure}
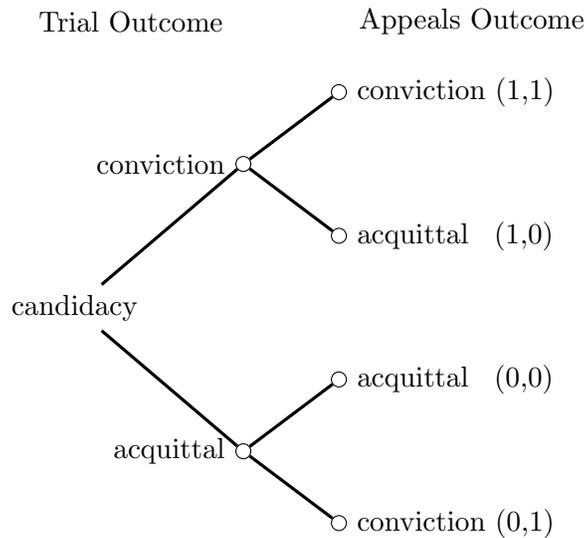

Since all criminal records are public, I can recover information disclosure effects for good and bad type politicians in groups (1) and (3). Using the institutional design described in section \ref{subsec:causal}, I implement an instrumental variables (IV) strategy where the effect of the disclosure of politician type is the local average treatment effects (LATE) for compliers, i.e., those who have been convicted or acquitted at both judicial review stages. I depict these groups in a classic IV framework in figure \ref{fig:ivtree}, where compliers are candidates with the same outcome at trial and on appeals.\footnote{Candidates who break the electoral code but are not detected are not part of this study, neither are candidates who have chosen not to appeal their trial sentence.}

Serving as a baseline for the IV strategy, I first estimate the following OLS model in three ways and using three different measures of electoral performance:
\begin{equation} \label{eq:reg1}
  \begin{split}
    y_{i} = \alpha + \rho \cdot c_{i} + X\beta + \sum \lambda_{i, k} + \varepsilon_{i}
  \end{split}
\end{equation}

The dependent variable $y_{i}$ measures are: (1) the probability of election, taking up value one when either the mayor or city council candidate had enough votes for election in their district; (2) the total vote share of candidate $i$ in their race; (3) the vote distance to the election cutoff, which is the percentage point margin between candidate $i$'s vote share and that of the single elected candidate (when running for mayor) or last elected candidate (when running for city council). Using outcome (1), I can measure the impact of crime disclosure on the most important outcome of any political campaign, i.e.~being elected; outcome (2) serves as a measure of the impact on candidate popularity; outcome (3) tells us about the relative benefit (or cost) of risking an electoral crime when candidates are trying to secure an electoral lead or narrow in on races in which they are trailing another candidate; $X$ is the matrix of candidate characteristics, such as candidate age, gender, marital and education status, political experience, and campaign spending; $\sum \lambda_{i, k}$ is a set of $k$ fixed-effects to capture any additional unobservable heterogeneity coming from party, election, and municipal factors shared by subsets of candidates.

The main independent variable is the binary indicator for convictions for electoral crime $c_{c}$ at the electoral court system for candidate $i$. These convictions are public information, so they are equivalent to the disclosure of criminal information to voters. If a candidacy has been rejected by the trial judge responsible for that electoral district, $c$ becomes one. I use convictions at trial in OLS regressions for benchmarking the biased effect on electoral performance; in reduced-form regressions, I replace convictions at trial for convictions on appeal -- which becomes one when the candidate has seen an unfavorable ruling at higher courts within the electoral system. The reduced-form regressions hint at any potential correlation between instruments and outcomes beyond the channel via the endogenous decision at trial (discussed in section \ref{sec:exclusionrestriction_paper1}). I lastly estimate model \refp{eq:reg1} using two-stage least squares (2SLS) regressions, in which I instrument convictions at trial for convictions on appeal. Since I am looking at appellate court decisions issued after election day, the exclusion restriction is straightforward as I measure the instrument \emph{after} observing the outcomes.\footnote{In addition to the temporal effect, the other theoretical arguments discussed in section \ref{sec:background_paper1} support the exogeneity of the instrument. Electoral judges are tenured state judges which have no ties to local politicians. Their wages, career prospects, and time on electoral bench are all independent of the action of mayors and city councilors.} Any effect of disclosing appellate decisions influences electoral performance only via their relationship with convictions at trial. I address additional concerns on violations to the exclusion restriction in the following sections, but the instrumental variables and the first-stage regression equations are:
\begin{equation} \label{eq:reg2}
  \begin{split}
    y_{i} = \alpha + \rho \cdot \hat{c}_{i, \text{trial}} + X\beta + \sum \lambda_{i, k} + \varepsilon_{i}
  \end{split}
\end{equation}
\begin{equation} \label{eq:regfirststage}
  \begin{split}
    c_{i, \text{trial}} = \alpha + \rho \cdot c_{i, \text{appeals}} + X\beta + \sum \lambda_{i, k} + \varepsilon_{i}
  \end{split}
\end{equation}

For every specification of equations \refp{eq:reg2} and \refp{eq:regfirststage}, I estimate versions excluding and including individual characteristics (matrix $X$) and fixed-effects $\sum \lambda_{i, k}$. In addition to instrument validity tests, I also report coefficient stability tests across different specifications to demonstrate that selection on unobservables is not driving the results, as discussed in \citet{AltonjiSelectionObservedUnobserved2005,NunnSlaveTradeOrigins2011,OsterUnobservableSelectionCoefficient2019,PeiPoorlyMeasuredConfounders2019}. I discuss and test other alternative, confounding explanations in the following sections and provide the empirical strategy at each stage of analysis.

\subsection{Inclusion Restriction Checks}\label{subsec:instrumentstrength_paper1}

The first step in this analysis is guaranteeing I have a strong instrument for the endogenous regressor of interest (conviction at trial). Table \ref{tab:reversals} provides us anecdotal evidence on the relationship between convictions at either stage of the judicial review process. The overall reversal rate of trial decisions is 18.58 percent. Reversals come mostly from candidacy cases that had been denied by trial judges (23.05 percent). The unconditional Pearson correlation coefficient between convictions at trial and on appeals is .637. These results make intuitive sense given the presumed quality of judges and standard sentencing (both in substance and form) discussed in previous sections.

A more robust test, however, is reported in table \ref{tab:firststage}. I present three first-stage regressions on the relationship between the endogenous variable (convictions at trial) and instrument (convictions on appeals). Across models progressively including candidate characteristics and municipal, electoral, and party fixed-effects, the coefficient on the instrument is always statistically significant (\emph{p}-value $<$ .01). The magnitude remains stable within the .629-.522 range, which means that a conviction on appeals explains roughly sixty percent of the outcome at trial. The positive relationship confirms the anecdotal evidence in table \ref{tab:reversals}.

\begin{figure}[!htbp]
\centering
\caption[figure2]{Instrument Point Estimates and CIs}\label{fig:firststage}
\vspace{-10pt}
\includegraphics[scale = .50]{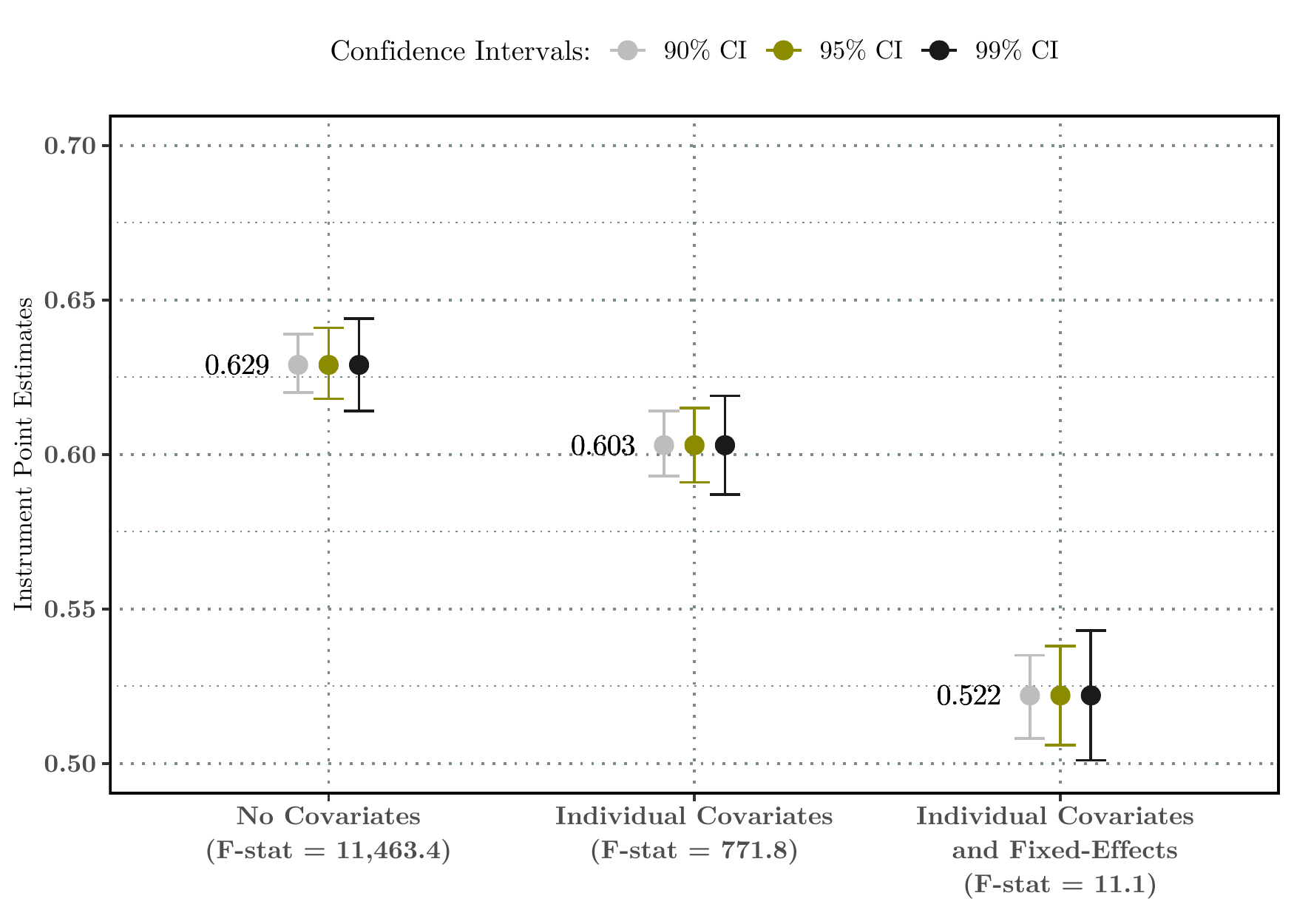}
\end{figure}

I additionally report each coefficient point estimate, confidence intervals (CIs), and \emph{F}-statistics for all three regressions in figure \ref{fig:firststage}. The inclusion of covariates and fixed-effects across models marginally shifts down the magnitude of instrument estimates. In all cases, however, the \emph{F}-statistic of excluded instruments remains greater than industry standards  at \emph{F} = 10 \citep{BoundProblemsInstrumentalVariables1995b}. It means that the first-stage model is significantly predicting the candidacy outcome at trial and confidently partials out the causal effect of convictions on electoral performance.

In table \ref{tab:hausman}, I present the Hausman tests for OLS consistency. I report the results for bivariate regressions between convictions at trial and on appeals for all outcomes.\footnote{I also run multivariate versions of Hausman tests, but there are no changes to \emph{p}-values. Results are available upon request.} Each row contains the \emph{F}-stat and \emph{p}-values for the null of OLS consistency. I reject consistency for outcomes one and two (\emph{p}-value $<$ .01) when using the full sample and for outcome three when splitting the sample into city council and mayor candidates (also \emph{p}-value $<$ .01). Since the vote distance to election cutoff is much smaller when votes are spread out across many candidates in proportional elections (city council) than in majoritarian elections (mayor), it makes sense to split the sample to analyze this performance measurement separately.

These tests confirm instrument choice and substantially support inclusion restriction conditions for causal identification under an IV design. After the results section, I also conduct exclusion restriction tests to provide further support for the information effects in this paper.

\section{Results}\label{sec:results_paper1}

Table \ref{tab:secondstageoutcome1} reports the effect of criminal information disclousre (via trial conviction) on the probability of election of each politician. For mayor candidates, this variable turns on when the candidate was the most voted in their election. For city council candidates, this variable turns on when the candidate has received enough votes to finish the election within the number of open seats in their municipality. For instance, if a municipality has 12 seats in its city council, a candidate who received the same number, or more, votes than the 12\textsuperscript{th} placed candidate has outcome value one.\footnote{City council elections are not necessarily decided in such manner; TSE tallies up all votes in a single election and divides them up by the number of seats available. All candidates who have more votes than this mark are automatically elected to office; remaining seats go to the coalitions who have rounded up more votes. Only rarely, however, all city councilors are elected like so. In most cases, votes are usually spread out across many candidates and coalitions, so being voted in as the last candidate within the number of available seats does guarantee their election and supports their coalitions to get further seats. In addition, this is a less strict way to define who is elected to city council such that, even if there are measurement errors in coding this outcome, the correct measurement would decrease the number of elected candidates and reinforce the conviction effect.} It is the most important outcome and directly tests the first theoretical claim suggested in section \ref{sec:theory_paper1}, that is, voters would impose electoral penalties using information cues about criminal activity; another way to state this is that convictions for electoral crimes signal candidate type and influence how voting decisions.

In columns 1-3 of table \ref{tab:secondstageoutcome1}, I report the OLS estimates of the effect of information shortcuts via convictions at trial. The point estimates start at a 9.3 percentage point reduction on the probability of election but decrease to 6.3 percentage points in model 3, which includes candidate controls and fixed-effects. All effects are significant (\emph{p}-value $<$ .01). Therefore, regardless of the specification, there is a negative baseline relationship between revealing criminal information and performance. Unsurprisingly, the inclusion of covariates and fixed-effects soaks up some of the variation in the conviction variable, and controls for observed factors potentially correlated with convictions.

This biased result alone is interesting. It suggests that judicial decisions are also relevant information mechanisms influencing voter decision, a relationship yet undocumented in the literature. In \citet{NicholsonInformationCuesRational2019}, there are many examples of information cues in the form of party labels, sociodemographic characteristics, politician performance, interest groups, or the media, but there are no examples of legal cues from court documents. This finding fills in an important gap in the literature by documenting the effect of disclosing information from a trustworthy source on voter behavior: in the U.S., people trust the judicial branch (68 percent) more than the executive (45 percent) and the media (41 percent) \citep{GallupGallupPollSocial2017,GallupGallupPollSocial2018}; in Brazil, they trust the judiciary (24 percent) more than political parties (7 percent) and the executive (6 percent) \citep{RamosRelatorioICJBrasil1o2017}. The unbiased effect of judicial information, however, is harder to assess. First, though voters trust the judiciary more than other sources of information, the vocabulary and intricacies of legal cases might make it harder to interpret judicial decisions. For instance, voters might not understand the different electoral crimes and assign equal punishment to all convictions. Second, the information released in court documents is a function of the conditions of individual electoral races. Repeated candidates, or incumbents, are more exposed to legal action than new candidates. I explore this latter mechanism first, and address the interpretation of court documents in section \ref{sec:hte_paper1}.

A plausible hypothesis here is that some electoral races are more relevant than others and, as such, there is more competition for seats than otherwise. Candidates might even be less likely to play by the rules and bring many unfounded claims against their opponents in an attempt to disqualify them from the race. Alternatively, some races might have more skilled politicians than others. In any of these cases, the resulting claims against candidates are not randomly distributed, and are likely correlated with unobservable factors at each race. To partial out such confounding effects, I implement the instrumental variables strategy discussed in section \ref{sec:methods_paper1}, and report its results in columns 4-6. Note that all IV coefficients have significantly larger magnitudes than their OLS equivalents (again at the one percent level). They range from $-$14.3 to $-$10.3 percentage points in models 4 and 6, respectively. They suggest an upward bias in OLS estimates of about 4 percentage points if we compare paired models; OLS predicts a smaller, weaker impact of information on performance. Along with the evidence of Hausman tests in section \ref{sec:methods_paper1}, these factors support IV consistency and its asymptotic convergence to the true, unbiased informational effect on voter response. For any given candidate, a conviction at trial alone would reduce their probability of election by 10.3 percentage points, according to my preferred model (column 6).

This result supports the negative information effect (suggested in section \ref{sec:theory_paper1}) and aligns with similar evidence in the literature. \citet{FerrazExposingCorruptPoliticians2008b} report a smaller effect of 7 percentage points for mayors when audit reports reveal corruption before elections in 2004. Though the effect here is larger for a less severe crime, the candidates in \citet{FerrazExposingCorruptPoliticians2008b}'s sample are generally much more experienced than in this paper. The share of reelected mayors in \citet{FerrazExposingCorruptPoliticians2008b} is 58.5, compared to 19.4 percent of experienced politicians in this sample, anecdotally suggesting that ability would indeed offset some of the negative information effect \citep[see][]{WintersLackingInformationCondoning2013,PereiraReelectingCorruptIncumbents2015}.

In table \ref{tab:secondstageoutcome2}, I report the results of the same regressions but on the vote share outcome. The OLS estimates are in columns 1-3 and show a negative and significant effect of criminal information on candidates' vote share, ranging from $-$9.7 to $-$5.5 percentage points. The IV effect is about five percentage points smaller than the OLS's. In my preferred model (column 6), the conviction effect significantly reduces vote share by 12.9 percentage points (\emph{p}-value $<$ .01). I should point out that the difference between OLS and IV parameters in this model is half the magnitude in the model for outcome one, which could create some skepticism about the marginal gain of using IV. The closer their magnitudes are, the smaller is the gain from trading bias (OLS) for consistency (IV). However, both the Hausman tests in section \ref{subsec:instrumentstrength_paper1} and the fact that the 99 percent CIs around OLS and IV coefficients never overlap indicate IV as the best choice for measuring the information effect of electoral crimes.

The effect size on outcome two is larger than in comparable studies. \citet{FerrazExposingCorruptPoliticians2008b} report a 10.4 percentage point decrease in vote share when mayors are running for reelection and have had corruption evidence released to the public prior to municipal elections in 2004. \citet{ChongDoesCorruptionInformation2015} run an experiment before the municipal elections in three Mexican states in 2009 and find a 1.1 decrease in incumbent mayors' vote share when corruption information is revealed to the public. The differences in research design, however, explain why the effect is smaller in these studies. First, both \citet{FerrazExposingCorruptPoliticians2008b} and \citet{ChongDoesCorruptionInformation2015} are looking at the effect for incumbent politicians when there is evidence of corruption. These politicians are likely more skilled than the average and thus offset the negative impact of information with their ability. Second, they also only look at mayors, rather than city councilors. The former have more visibility in local politics than the latter, which again could offset the negative impact of criminal information. When I reestimate the model in column 6 for the sample of mayor candidates who have political experience (unreported here but available upon request), the information effect remains significant and negative but falls to 1.4 percentage points -- marginally greater than \citet{ChongDoesCorruptionInformation2015}. Thus, I have reason to believe the effect size here is consistent with a sample of less skilled local politicians.

I lastly investigate the information effect for outcome three, vote distance to election cutoff. This effect represents how much information disclosure helped getting away (or closer) to the number of votes needed for election. In this analysis, I split the sample into city council and mayor candidates because of the meaningful differences in each office race. Mayor elections follow majority rule; city council elections follow proportional rule. As such, the number of candidates is much smaller, and the votes needed for election much larger, in mayor elections. Therefore, the distance to election is not uniform across office type; in other words, a one percentage point distance is much harder to come by in city council rather than mayor races.

Table \ref{tab:secondstageoutcome3} presents these results. Columns 1-2 display OLS specifications and columns 3-4 display IV models. I only report regressions with individual controls and fixed-effects. I find that releasing information of conviction at trial has again a negative and significant effect (at the one percent level) on the vote distance to the election cutoff across all models. For the city council sample, the IV coefficient points to 0.9 percentage point less in the distance to election than in the absence of a crime; for the mayor sample, this effect is 14.1 percentage points. Thus, candidates accused, and found guilty, of violating electoral law generally place further away from the necessary votes to guarantee election -- in line with the impact on outcomes one and two.

I document a robust, negative criminal information effect on electoral performance across all models. The IV estimates never overlap with their OLS equivalents. When different research designs are accounted for, these results align well with previous evidence in the literature. In the following sections, I conduct multiple robustness checks to support the negative, unbiased, and significant effect of judicial information.

\section{Exclusion Restriction Checks}\label{sec:exclusionrestriction_paper1}

In section \ref{subsec:instrumentstrength_paper1}, I carried out three tests validating the inclusion restriction of convictions on appeal as an instrument for convictions at trial. In section \ref{sec:background_paper1}, I also discussed how the TSE's appellate process qualifies for the exclusion restriction. Though there are no empirical tests for the exclusion restriction, I conduct two indirect checks which support my instrument choice: (i) correlation tests including the instrument in the first-stage; and (ii) coefficient stability tests following \citet{AltonjiSelectionObservedUnobserved2005}, \citet{OsterUnobservableSelectionCoefficient2019}, and \citet{PeiPoorlyMeasuredConfounders2019}.

\subsection{Instrument Included in Second-Stage}\label{subsec:instrumentss_paper1}

\begin{figure}[!htbp]
\centering
\caption[figure3]{Instrument Correlation with Covariates}\label{fig:instrumentcorrelation}
\vspace{-10pt}
\includegraphics[scale = .65]{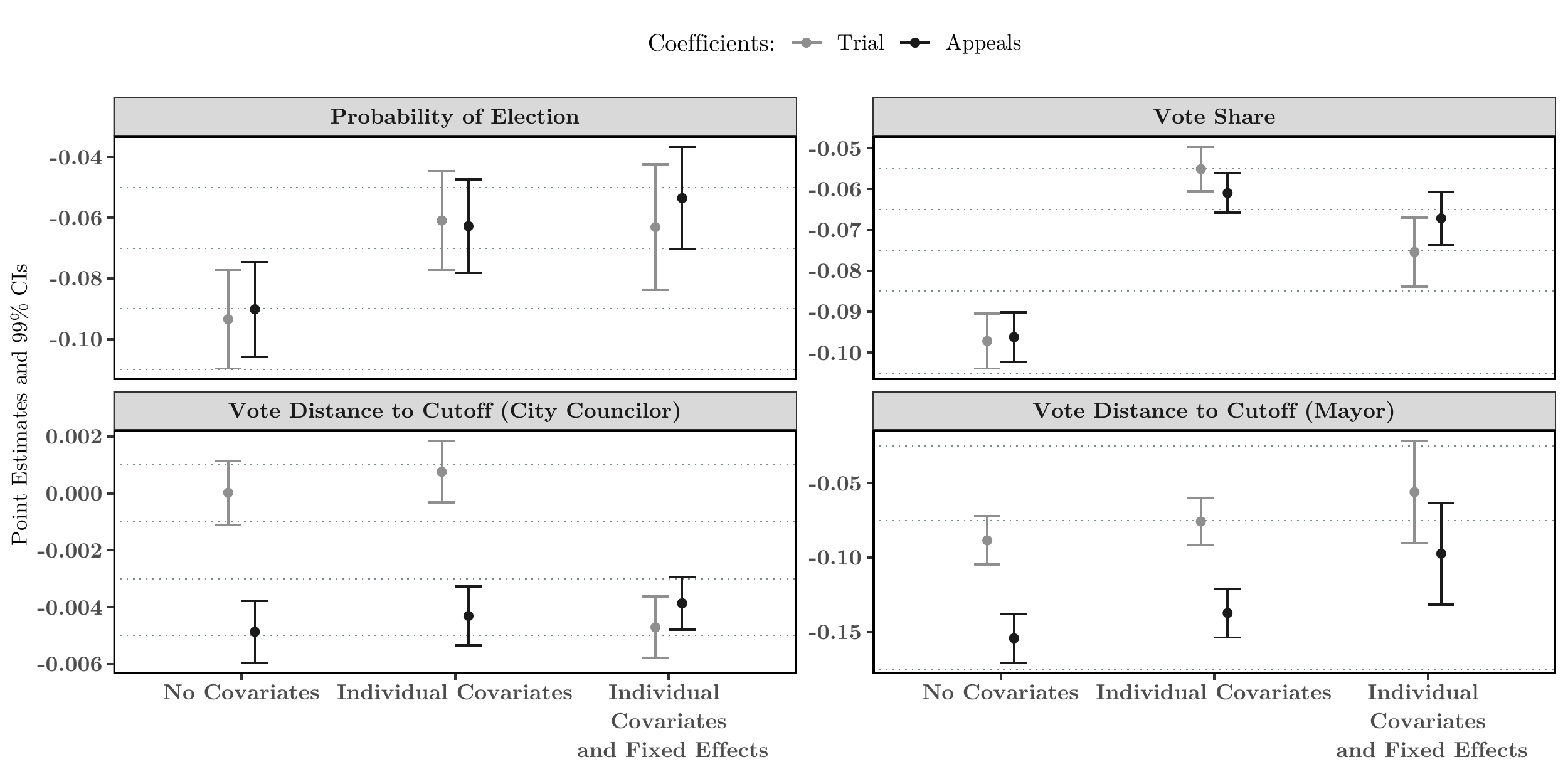}
\end{figure}

The first test in support of the exclusion restriction is relatively straightforward. I run two sets of regressions for all outcomes. In group one, the main independent variable is the conviction at trial. In group two, it is the conviction on appeals. For each set, I estimate a bivariate model, a multivariate model including individual controls, and a multivariate model including individual controls and all fixed-effects (party, municipal, and election effects). The regressions in group one are the same as the OLS regressions in table \ref{tab:secondstageoutcome1}. The regressions in group two just replace the endogenous variable for the instrument. The latter reduced-form approach is only recommended if I am able to reject OLS consistency in favor of IV -- which is the case here according to the Hausman test in table \ref{tab:hausman}. I report the results in figure \ref{fig:instrumentcorrelation}, where there are four panels, one for each of the outcomes discussed in section \ref{sec:results_paper1}. In light gray, I plot the OLS coefficient on convictions at trial and its 99 percent CI for all outcomes and regression models, yielding a total of 12 estimates and CIs. I do the same for convictions on appeals, the reduced-form OLS model, and produce the same estimates and CIs (in black).

It is clear that the instrument is significantly correlated with the outcome of interest. No 99 percent CI in the appeals regressions includes zero. This is expected when the instrument passes inclusion restriction tests. The more important result, however, is the similarity of point estimates using either trial and appeals convictions. For all models including individual covariates and fixed-effects, all point estimates of appeals coefficients fall inside trial CIs. The correlation between either variable, covariates, and fixed-effects is the same, meaning that the instrument is not adding any more variance to the trial regression. In other words, there is additional support for $cov(z, X) = cov(x, X)$. Almost all of the effect of instrument $z$ on outcome $y$ occurs via its correlation with $x$. Along with the evidence in the previous section and the institutional design of judicial review of candidacy registration cases, I can confidently say that there is no independent effect of the instrument on the outcomes of interest.

\subsection{Coefficient Stability Tests}\label{subsec:coefstab_paper1}

The most common way to address omitted variable bias is to include controls in the regression of interest. In this paper, I repeatedly report parameter estimates progressively including candidate controls, party, municipal, and election fixed-effects. In many cases, however, the set of controls does not fully identify confounding effects. In fact, scholars rarely use the full set of confounding factors; instead, they use the \emph{observed}, available confounders. Unless available variables fully capture the confounding set, selection on unobservables could still explain a significant portion of the parameters we are estimating in linear models.

\citet{OsterUnobservableSelectionCoefficient2019} formalizes this point. She suggests that coefficient stability across regression models is only a reliable indication of unbiasedness if scaled by changes in the amount of regression variation explained by independent variables. In other words, the coefficient of interest should move relatively less than $R^{2}$, indicating the stability of effect size as the researcher shifts explained variation from the error term to the matrix of independent variables. The following equation in \citet{OsterUnobservableSelectionCoefficient2019} translates this idea:
\begin{equation} \label{eq:coef1}
\beta^{*} = \tilde{\beta} - \delta \cdot [\beta^{0} - \tilde{\beta}] \cdot \frac{R_\text{max} - \tilde{R}}{\tilde{R} - R^{0}}
\end{equation}

Where $\beta^{*}$ is the bias-adjusted coefficient; $\tilde{\beta}$ is the coefficient in the unrestricted regression; $\beta^{0}$ is the coefficient in the restricted regression; $\tilde{R}$ and $R^{*}$ are their respective $R^{2}$. In this setting, we are interested in adjusting $\delta$ and $R_\text{max}$ such that we can test how $\tilde{\beta}$ fares against $\beta^{*}$: $\delta$ is the degree of selection on unobservables (i.e., proportion of outcome variation explained by unobservable variation over observable variation); $R_\text{max}$ is the theoretical population variance explaining the outcome. \citet{AltonjiSelectionObservedUnobserved2005} and \citet{OsterUnobservableSelectionCoefficient2019} suggest that $\delta = 1$ is a  reasonable threshold for coefficient stability, which means that unobservable and observed variables are equally able to explain $\tilde{\beta}$.

To support the exclusion restriction, I use equation \refp{eq:coef1} to compare the effect of coefficient magnitudes on the probability of election using the two conviction variables (at trial and on appeals). It is a similar exercise to the test in section \ref{subsec:instrumentss_paper1}, where I run two sets of regressions one with each conviction variable. I generate the bias-adjusted coefficients in each set, $\beta_\text{trial}^{*}$ and $\beta_\text{appeals}^{*}$, under different $R_\text{max}$ assumptions to verify whether coefficient distributions overlap across models. In \citet{OsterUnobservableSelectionCoefficient2019}, the author suggests that $[\tilde{\beta}, \beta^{*}]$ represents bounds on true value of coefficients when adding controls to regression models. If these confidence intervals overlap, I have additional evidence that $cov(z, X) = cov(x, X)$ such that unobservable variation would not be driving the magnitude of the main results in IV models.

I present the results in table \ref{tab:coefstability}. It depicts the bounded estimates suggested by \citet{OsterUnobservableSelectionCoefficient2019} under different $R_\text{max}$ assumptions and $\delta = 1$. It indicates the potential bias in coefficient estimates by looking at how their magnitudes would change if I was able to use more of the variation in the error term to explain the outcome variable. In other words, I compare potential coefficient estimates if I could fully identify all omitted variables and included them in the model, shifting unexplained variation to explained variation. Columns (1) and (2) show that the bounded estimates of either coefficient magnitude overlap, suggesting that there is no independent correlation of the error term with the instrument. In other words, $cov(z | x, \varepsilon) = 0$.

\section{Alternative Explanations}\label{sec:alternatives_paper1}

There are additional threats to validity beyond the inclusion and exclusion restriction. They originate from data-generation processes and cannot be detected by any single identification strategy. In this section, I explore how strategic changes of judge sentencing behavior, voter engagement, or candidate campaigning behavior during the judicial review process could explain the information effect. If they are true, then what I am picking up is something different than voter punishment for candidate's bad type.

\subsection{Heterogeneous Sentencing across Review Stages}\label{subsec:heterosentencing_paper1}

The first alternative explanation to the criminal information effect is the potential change in judges' sentencing behavior over case duration. Both elections and judicial review coincide in time, and judges could change how they sentence candidates based on campaign promises, policy positions, or even as a response to electoral results: judges might have a hard time issuing sentences preventing candidates who received the most votes from taking up office. To test this mechanism, I provide indication that judicial decision-making factors are not differentially affecting trial and appeals sentencing, i.e., that conviction signals are the same over the electoral period. Judges should be using the same criteria, and weighing them the same, when reviewing candidates' cases at trial and on appeal. To that end, I implement a modified version of \citet{PeiPoorlyMeasuredConfounders2019}'s covariate balancing test, which constitutes in regressing variables of interest on other covariates.

The test is as follows. First, I run two independent regressions with conviction variables on the left-hand side. The respective dependent variables are conviction at trial and on appeals. I report these regressions in columns 1 and 2 of table \ref{tab:heterogeneoussentencing}. Besides including the same covariates as before, I also include the main electoral outcome, whether the candidate had enough votes to take up office, on the right-hand side. It is the most important sentencing factor, as judges might be less willing to convict candidates once they know these people have had enough votes to take up office. I also include party-fixed effects, and cluster standard errors at the municipality-election pair level to account for the shared variation of standard errors at the judge level (each judge oversees one electoral district one election at a time). Next, I run a \emph{t}-test on the difference between each parameter across regressions. Columns 3-6 respectively report the difference in coefficients, the joint distribution of standard errors, \emph{t}-stats, and the \emph{p}-value of each test. The null hypothesis is that the parameter difference is zero, meaning that the factor has the same effect on the trial and the appeals decision.

The results in table \ref{tab:heterogeneoussentencing} are evidence of homogeneous sentencing over the judicial review process. The difference in parameters is not statistically significant at any industry standard. No \emph{p}-value is smaller than .13 (column 6). Moreover, it does not seem that being elected to office (outcome one) changes the way judges rule on a particular candidate's case; the difference of $-$.024 is not significant (\emph{p}-value = .649). This is strong evidence in favor of homogeneous sentencing, as judges do not seem to be changing their sentencing behavior over time. This is consistent with the institutional design of the electoral court system in Brazil. In local elections, trial rulings are issued by electoral district judges, who face both career and monetary incentives independent of local politics; appeals are decided by a panel of three judges at the state level, who are appointed to the electoral bench by fellow judges in state and federal systems, and the President of Brazil. Therefore, there is no evidence that heterogeneous sentencing would be driving the effect discussed in section \ref{sec:results_paper1}.

\subsection{Voter Disengagement Effect}\label{subsec:voterbehavior_paper1}

A second source of concern is whether voters also change their behavior once they learn candidates' trial outcomes. While this could mean they change their votes for someone else (my hypothesis), this could also mean disengagement of the political process altogether. In this scenario, rather than punishing candidates for criminal behavior, voters would become frustrated with politics. The mechanism behind the information effect would be disengagement rather than punishment \citep{PavaoCorruptionOnlyOption2018,ChongDoesCorruptionInformation2015}.

I cannot disentangle this effect by only looking at the main results from section \ref{sec:results_paper1}. Instead, I have to look at other election outcomes to check for signals of disengagement. The first signal comes from voter turnout. If voters are frustrated with the electoral process, e.g., they believe candidates are dishonest because they observe convictions for electoral crimes, then one potential reaction is simply skip voting. Though Brazil has mandatory voting in place, the costs of not voting are negligible. Voters only have to fill out a no-show form, either online or in person. If they do not, they have to pay a \$1 fine. In this case, a decrease in voter turnout if the TSE convicts more candidates would be evidence of disengagement. Another related signal is the number of invalid votes in each municipal election. Voters could show up to the ballot but intentionally cast a blank vote or type in a non-existing candidate number in the electronic voting machine, both qualifying as invalid (spoiled) votes in Brazil. Thus, another evidence of disengagement could come from a higher number of invalid votes when there are more convicted candidates running for office.

I report the results of these tests in table \ref{tab:voterbehavior}, where I aggregate up voter turnout and invalid votes to the party and election-level.
The most accurate aggregation of convictions would be the computation of the share of invalid candidacies over the total number of candidacies at each race. If there were 100 candidates at a single race and 25 of these individuals were convicted at trial, the share of invalid candidacies would be 25 percent. However, the proportion of invalid candidacies over total candidacies is very small. Out of an initial sample of $\sim$200,000 local office candidates, only 16,791 people have unfavorable decisions at trial but stay in the race. This means that the share of invalid candidacies over total candidacies would have a very small variance across all elections in my sample. It would be hard to reject the null between share of invalid candidacies and voting outcomes. I could mistake the null effect coming from low variation for the null effect of no correlation. Instead, I want to make it easy to reject the null, for then I am more certain that there is no voter disengagement effect. Therefore, I construct measures of the share of invalid candidacies, both at trial and on appeals, \emph{over the total number of office vacancies in every election}.\footnote{In mayor races, for instance, this means that the share of invalid candidacies might yield values greater than one for the simple reason that there is always just one open spot for mayor in each municipality. These tests place a higher bar for rejecting the voter disengagement hypothesis, providing yet more confidence in my results.} I intentionally inflate the denominator of the share of invalid candidacies in the regressions for each disengagement outcome and aggregation level. In columns 1 and 3, I find no relationship between convictions and voter turnout for both aggregation levels. This means that the number of invalid candidacies per party or election does not affect turnout. Voters still cast their ballots regardless of the number of invalid candidacies. In columns 2 and 4, the share of invalid candidacies has a significant and positive relationship with invalid votes. Voters still get out to vote, but once at the voting station, they cast more invalid votes when there are more convicted candidates (either per party or election). However, the effect size is almost meaningless. A one percentage point increase in the share of invalid candidacies by party only increases invalid votes by 0.276 percent; similarly, one percentage point increase in the share of invalid candidacies by election only increases invalid votes by 0.183 percent. Together, these results do not point to disengagement as the main driver behind the criminal information effect.

\subsection{Candidates Quit Campaigning}\label{subsec:candidatebehavior_paper1}

The third alternative explanation would come from campaign responses after candidates receive an unfavorable trial ruling. Candidates who receive such ruling might anticipate the eventual disqualification of their candidacy by the appellate panel such that they, partially or entirely, quit campaigning. As a consequence, the hit to electoral performance would come from a candidate effort rather than a voter punishment effect. I believe this to be a minor problem for the simple reason that my sample only contains candidates who remained engaged in their race to office -- evidenced by their filing of an appeal against their trial sentence. This is a limitation of the instrumental variables design more generally, and it is likely that the candidates in this sample are not the same as the population of candidates in local elections in Brazil, but it stills fends against claims of candidate disengagement. Second, any strategic candidate believing they have a shot at election would do well to keep campaigning since the judicial penalty (dismissal of candidacy) is small compared to the benefit of holding office. Candidates would be willing to take a gamble with small risk but high reward.

Nevertheless, I provide anecdotal evidence to dismiss concerns about candidates making such large shifts in their campaign strategies. In table \ref{tab:candidatebehavior}, I compare the mean campaign expenditure across various subgroups of candidates. Campaign expenditures do not fully capture the extent to which campaigns are adjusted, but are a good proxy to understand campaigning behavior. I first compare mean spending for candidates by the type of trial and appeals ruling they receive. The mean spending of candidates with favorable rulings at trial (R\$79,099, or $\sim$\$18,800) is much higher than candidates with unfavorable rulings (R\$35,402, or $\sim$\$8,430). The same is true for outcomes on appeals. I can thus anecdotally claim that there is an association between campaign spending and judicial outcomes, which is not surprising; for precisely this reason, I control for campaign spending in all regressions of this paper. The more interesting result, however, is the bottom row of table \ref{tab:candidatebehavior}. For the subgroup of candidates who received an unfavorable ruling at trial, campaign spending is not associated with better outcomes on appeal. Campaign spending is statistically the same (\emph{p}-value = .53), and their respective means are R\$39,004 and R\$34,324. Unfortunately, I cannot observe expenditure dates, which would be a better proxy for campaign engagement, but the indirect evidence here is that candidates who needed to reverse a trial ruling have not differently spent money on their campaigns. If they had spent more, we would see an engagement effect trying to revert unfavorable rulings. If they had spent less, we would find evidence in favor of candidate disengagement. None of these explanations apply, and this is additional evidence that the main effect can be attributed to voter punishment.

\section{Heterogeneous Electoral Punishment}\label{sec:hte_paper1}

Besides the primary effect, I want to identify whether voters impose heterogeneous punishment by information type. This is called the \emph{crime type} effect. In an ideal world, severe electoral code violations should be met with harsher electoral penalties than trivial violations. While it makes sense to expect that punishment should fit the crime, the literature is filled with cases in which this is not true. Voters might not punish candidates as expected because they trade off crimes for public goods \citep{PereiraReelectingCorruptIncumbents2015}; they might not trust the source of information on criminal behavior, or might not understand the information being disclosed \citep{WintersLackingInformationCondoning2013,Weitz-ShapiroCanCitizensDiscern2017}; they might even lack options, and end up voting for the \emph{least} dishonest candidate \citep{PavaoCorruptionOnlyOption2018}. None of these studies, however, looks at the disclosure of legal information in the form of judicial decisions. This information might be more credible but also harder to understand.

To estimate such effect, I collect and code court documents for all candidates running for municipal office in 2012 and 2016, and a few in 2004 and 2008. For 2016, I also have official TSE classification of conviction types.\footnote{Though documents are public, the TSE had many errors in the system that prevented me from downloading almost all case files for 2004 and 2008.} There are eight conviction categories, and I group them in two classes: (1) violations to the Law of Elections (1995), the Law of Political Parties (1997), the Clean Records Act (2010), and (2) documentation problems. Severe rule-breaking are egregious cases in which either parties or candidates are in breach of more substantial elements of electoral law, such as using illegal campaign strategies, channeling slush funds for campaign ads, sexual assault, murder, corruption, and having previous outstanding judicial convictions preventing them from running for office. These crimes come from the legislation under the jurisdiction of the electoral court: the Electoral Code of 1965, the Law Establishing Conditions for Ineligibility to Public Office (1990), the Law of Political Parties (1995), the Law of Elections (1997), and the Clean Records Act of 2010. This is why I group the first three laws into the \emph{severe} rule-breaking class. Trivial rule-breaking are cases in which candidates lack a procedural requirement to run for office, such as missing documentation, not meeting deadlines, or not having submitted notarized documents. For instance, they could have forgotten to include a copy of their ID card in their application, or they could have missed a deadline in the application process. They make up the documentation group. Next, I use seven machine classification models to predict these two conviction classes for all candidates in the 2004, 2008, and 2012 elections. I pre-process the text of the sentences and convert all words to unigram and bigram combinations and use them as inputs in Linear-Support Vector, Multinomial Bayes, Logistic, Random Forest, Adaptive and Gradient Boosting classification models. I also convert unique words in sentences to  300-dimension word tensors using Facebook's FastText dataset \citep{JoulinBagTricksEfficient2016}. These tensors serve as weights in the last classification model using Deep Neural Networks. Once these classification algorithms have been trained on 2016 data, I apply the models to predict classes in earlier elections, \emph{which were not disclosed by the TSE}, in a standard application of supervised machine classification.\footnote{I discuss these classification algorithms  in more detail in appendix \ref{sec:machinelearning_paper1}.} These predictions form a new binary variable $b_i$, where $b_i = 1$ represents a severe violation and $b_i = 0$ represents a trivial violation of electoral law.

I test the crime type effect by including $b_{i}$ in equation \refp{eq:reg2}. The parameters describing each relevant effect are the conviction effect ($\rho_{0}$), the crime type effect ($\rho_{1}$), and their interaction ($\rho_{2}$). Equation \refp{eq:reg3} summarizes these relationships:
\begin{equation} \label{eq:reg3}
  \begin{split}
    y_{i} = \alpha + \rho_{0} \cdot \hat{c}_{i, \text{trial}} + \rho_{1} \cdot {b}_{i} + \rho_{2} \cdot \hat{c}_{i, \text{trial}} \times {b}_{i} + X\beta + \sum \lambda_{i, k} + \varepsilon_{i}
  \end{split}
\end{equation}

Note that, in the absence of the conviction at trial, $\rho_{1}$ is just the raw association between breaching electoral law and electoral performance, and represents the potential gain of engaging in an activity prohibited by law. To make the interpretations clearer, I lay out the four alternative explanations of the joint effect of conviction and crime type in table \ref{tab:htehypotheses}: if $\rho_{1} = 0$, engaging in an electoral crime has no electoral payoff; if $\rho_{1} > 0$, engaging in an electoral crime has a positive electoral payoff; if $\rho_{2} = 0$, voters punish candidates the same, regardless of the accusation against them; if $\rho_{2} < 0$, voters fit punishment to the crime; in other words, severe violations receive larger electoral penalties.

Table \ref{tab:hte} reports results of the heterogeneous information effect by crime type. The first result that stands out is that the criminal information effect $\rho_{0}$ remains significant and negative. When compared to the outcomes in the primary results section, one also notices that the magnitude falls by approximately as much as the electoral gain of severely breaching electoral law. For instance, in the main results, the probability of election falls by 10.3 percentage points when crimes are disclosed by the electoral court, which is close to the subtraction of electoral gain from the information effect in table \ref{tab:hte}: $-6.2 - 6.9 = -13.1$. The same is true for vote share: the main effect of $-$12.9 is close to the subtraction of $\rho_{1}$ from $\rho_{0}$: $-9.5 - 5.2 = -14.7$. These results provide support for the offsetting, positive electoral effect of substantially breaching electoral law. In addition, I reject the homogeneous punishment hypothesis. The $\rho_{2}$ effect is negative and significant in the model for outcomes one and two. This result means that voters punish candidates convicted for more severe violations of electoral law harsher than those who found guilty of trivial violations. The magnitude of punishment is close to 6 percentage points for the probability of election or vote share.

The most suited explanation to the heterogeneous effect is in the bottom right quadrant of table \ref{tab:htehypotheses}. There is an electoral gain of engaging in severe electoral rule-breaking, but voters punish these candidates more harshly when the judiciary makes the severe violation public. This is a particularly interesting result as it supplements the findings in \citet{Weitz-ShapiroCanCitizensDiscern2017}. Using a vignette experiment, the authors document a positive relationship between information credibility and intention to vote for mayor candidates in Brazil; here, I find that voters acted on criminal information in local elections in Brazil between 2004 and 2016.

The results for outcome three are, however, puzzling. I find no heterogeneous effect on vote distance to cutoff for the city council sample and a positive effect for the mayor sample. One interpretation is electoral crimes have different effects across the distribution of candidates. Voters might not care if candidates for city council seats have committed severe or trivial crimes because they have limited ability to implement local policies in addition or in lieu of mayors'. This would explain the null result in column 3 of table \ref{tab:hte}. For mayor candidates, it could be that voters interpret criminal records as the ability to get things done, as documented in \citet{WintersLackingInformationCondoning2013} and \citet{PereiraReelectingCorruptIncumbents2015}. In any case, this should be an important avenue for future research.

\section{Conclusion}\label{sec:conclusion_paper1}

This project aims at uncovering the effect of electoral crime disclosure on performance. Supplemental to existing literature looking at flagrant electoral fraud and electoral malfeasance in non or partially democratic regimes, I provide evidence on less known, less understood electoral practices that can also shape the results of elections. I also suggest that judicial information is an important factor for voter behavior in addition to other information shortcuts documented in the literature (party labels, performance, ethnicity, gender, interest groups, and the media, for example). I document this information effect in Brazil, one of the largest and highest quality electoral democracies in the world at the time of the study. Further contributions of this paper are the use of court documents as data and causal identification using the institutional design of judicial review in the Brazilian electoral court system.

I find substantial and significant negative effects of criminal information on electoral performance. Being convicted of an electoral crime reduces the probability of election for mayor and city council candidates in Brazil by 10.3 percentage points. It also reduces vote share by 12.9 percentage points. These results are robust to a series of checks on the inclusion or exclusion restriction, and to strategic changes of judge, voter, and candidate behavior. I further find that voters impose harsher electoral punishment conditional on the severity of the information disclosure. There is a positive, independent effect of engaging in severe rule-breaking on electoral performance, but once voters know about this, they hold politicians accountable at the ballot.

This study is relevant for multiple policy reasons. I offer additional evidence claiming the existence of a negative relationship between electoral crime information and electoral performance beyond just corruption. Knowing that voters punish bad behavior, skilled politicians and policymakers can increase monitoring, detection, and prosecution of electoral crimes as a means of weeding out low-quality office-seeking candidates. A final implication of this project is a discussion on the effectiveness of electoral oversight authorities in the first place, a feature shared by other developing countries such as Mexico, India, and South Africa. While on the one hand they might prevent low-quality candidates from running, and eventually, being elected, they might simultaneously create barriers to entry that are detrimental to political competition and to the democratic process in developing countries.

\pagebreak

\pagebreak

\section*{Tables and Figures} \label{sec:tables_paper1}

\begin{table}[!htbp]
\centering
\caption[table1]{Descriptive Statistics}\label{tab:sumstats}
\vspace{-10pt}

\centering
\scriptsize
\setlength{\tabcolsep}{-1pt}
\begin{tabular}{@{\extracolsep{10pt}}lcrrrr}
\\[-1.8ex]\hline
\hline \\[-1.8ex]
& \multicolumn{1}{c}{\emph{n}} & \multicolumn{1}{r}{Mean} & \multicolumn{1}{c}{St. Dev.} & \multicolumn{1}{c}{Min} & \multicolumn{1}{c}{Max} \\
\hline \\[-1.8ex]
Age                                        & 16,791 & 46.63 & 11.28 & 17 & 86 \\
Male                                       & 16,791 & .781 & .413 & 0 & 1 \\
Political Experience                       & 16,791 & .082 & .275 & 0 & 1 \\
Campaign Expenditures (in R\$)             & 16,791 & 49,924 & 458,081 & 0 & 4e+07 \\
Convicted at Trial                         & 16,791 & .577 & .494 & 0 & 1 \\
Convicted on Appeal                        & 16,791 & .497 & .500 & 0 & 1 \\
Probability of Election                    & 16,791 & .194 & .396 & 0 & 1 \\
Total Vote Share (in p.p.)                 & 16,791 & .082 & .161 & 0 & 1 \\
Vote Distance to Election Cutoff (in p.p.) & 16,791 & $-$.088 & .080 & $-$.500 & .500 \\
\hline \\[-1.8ex]
\hline
\end{tabular}

\end{table}

\begin{table}[!htbp]
\centering
\caption[table2]{Electoral Crime Rulings}\label{tab:reversals}
\vspace{-10pt}

\centering
\scriptsize
\begin{tabular}{@{\extracolsep{12pt}}rrrr}
\\[-1.8ex]\hline
\hline \\[-1.8ex]
& \multicolumn{2}{c}{\emph{Appeals}} & Percent \T \B \\
\multicolumn{1}{r}{\emph{Trial}} & Affirmed & Reversed & Reversed \T \B \\
\hline \\[-1.8ex]
Not Convicted & 6210 & 885  & 12.47 \T \B \\
Convicted     & 7461 & 2235 & 23.05 \T \B \\
\hline \\[-1.8ex]
\hline
\end{tabular}

\end{table}

\begin{table}[!htbp]
\centering
\caption[table3]{First-Stage Regressions}\label{tab:firststage}
\vspace{-10pt}

\centering
\scriptsize
\setlength{\tabcolsep}{-2pt}
\begin{tabular}{@{\extracolsep{4pt}}lD{.}{.}{-3} D{.}{.}{-3} D{.}{.}{-3} }
\\[-1.8ex]\hline
\hline \\[-1.8ex]
& \multicolumn{3}{c}{Outcome: Convicted at Trial} \\
\cline{2-4}
\\[-1.8ex] & \multicolumn{1}{c}{(1)} & \multicolumn{1}{c}{(2)} & \multicolumn{1}{c}{(3)}\\
\hline \\[-1.8ex]
 Convicted on Appeal & .629^{***} & .603^{***} & .522^{***} \\
                     & (.006) & (.006) & (.008) \\
                     & & & \\
\hline \\[-1.8ex]
Individual Controls & \multicolumn{1}{c}{-} & \multicolumn{1}{c}{Yes} & \multicolumn{1}{c}{Yes} \\
Fixed-Effects & \multicolumn{1}{c}{-} & \multicolumn{1}{c}{-} & \multicolumn{1}{c}{Yes} \\
\hline \\[-1.8ex]
Observations & \multicolumn{1}{c}{16,791} & \multicolumn{1}{c}{16,791} & \multicolumn{1}{c}{16,791} \\
\textit{F}-stat & \multicolumn{1}{c}{11,463.4$^{***}$} & \multicolumn{1}{c}{771.81$^{***}$} & \multicolumn{1}{c}{11.10$^{***}$} \\
\hline
\hline \\[-1.8ex]
\multicolumn{4}{p{8.5cm}}{\textit{Note:} First-Stage regressions here report the correlation between being convicted at trial and being convicted on appeal for all candidates who have had their candidacy challenged under charges of electoral irregularities. I present results including and excluding individual politician characteristics; municipal, electoral, and party fixed-effects; and use robust standard errors. $^{*}$p$<$0.1; $^{**}$p$<$0.05; $^{***}$p$<$0.01} \\
\end{tabular}

\end{table}

\begin{table}[htbp]
\centering
\caption[table4]{Hausman Test of Instrument Strength}\label{tab:hausman}
\vspace{-10pt}

\centering
\scriptsize
\setlength{\tabcolsep}{0pt}
\begin{tabular}{@{\extracolsep{4pt}}lD{.}{.}{-2}D{.}{.}{-3}r}
\\[-1.8ex]\hline
\hline \\[-1.8ex]
\multicolumn{1}{l}{Outcome} & \multicolumn{1}{r}{Hausman Statistic} & \multicolumn{1}{r}{\textit{p}-value} \T \B & Sample  \\
\hline \\ [-1.8ex]
1. Probability of Election           & 45.03  & .000 & Full \\
2. Total Vote Share                  & 374.99 & .000 & Full \\
3. Vote Distance to Election Cutoff  & 253.02 & .000 & Full \\
% \hspace{10pt}3.1. City Councilor & 211.5  & .000 & Split \\
% \hspace{10pt}3.2. Mayor          & 449.45  & .000 & Split \\
\hline \\[-1.8ex]
\hline
\end{tabular}

\end{table}

\begin{table}[!htbp]
\centering
\caption[table5]{The Effect of Electoral Crime on the Probability of Election}\label{tab:secondstageoutcome1}
\vspace{-10pt}

\centering
\scriptsize
\setlength{\tabcolsep}{-2pt}
\begin{tabular}{lD{.}{.}{-3}D{.}{.}{-3} D{.}{.}{-3} D{.}{.}{-3} D{.}{.}{-3} D{.}{.}{-3} }
\\[-1.8ex]\hline
\hline \\[-1.8ex]
 & \multicolumn{6}{c}{Outcome: Probability of Election} \\
\cline{2-7}
\\[-1.8ex]
 & \multicolumn{1}{c}{OLS} & \multicolumn{1}{c}{OLS} & \multicolumn{1}{c}{OLS} & \multicolumn{1}{c}{IV} & \multicolumn{1}{c}{IV} & \multicolumn{1}{c}{IV} \\
\\[-1.8ex] & \multicolumn{1}{c}{(1)} & \multicolumn{1}{c}{(2)} & \multicolumn{1}{c}{(3)} & \multicolumn{1}{c}{(4)} & \multicolumn{1}{c}{(5)} & \multicolumn{1}{c}{(6)}\\
\hline \\[-1.8ex]
 Convicted at Trial & -.093^{***} & -.061^{***} & -.063^{***} & -.143^{***} & -.104^{***} & -.103^{***} \\
                    & (.006) & (.006) & (.008) & (.010) & (.010) & (.013) \\
                    & & & & & & \\
\hline \\[-1.8ex]
Individual Controls & \multicolumn{1}{c}{-} & \multicolumn{1}{c}{Yes} & \multicolumn{1}{c}{Yes} & \multicolumn{1}{c}{-} & \multicolumn{1}{c}{Yes} & \multicolumn{1}{c}{Yes} \\
Fixed-Effects & \multicolumn{1}{c}{-} & \multicolumn{1}{c}{-} & \multicolumn{1}{c}{Yes} & \multicolumn{1}{c}{-} & \multicolumn{1}{c}{-} & \multicolumn{1}{c}{Yes} \\
\hline \\[-1.8ex]
Observations & \multicolumn{1}{c}{16,791} & \multicolumn{1}{c}{16,791} & \multicolumn{1}{c}{16,791} & \multicolumn{1}{c}{16,791} & \multicolumn{1}{c}{16,791} & \multicolumn{1}{c}{16,791} \\
\textit{F}-stat & \multicolumn{1}{c}{231.52$^{***}$} & \multicolumn{1}{c}{45.18$^{***}$} & \multicolumn{1}{c}{5.99$^{***}$} & \multicolumn{1}{c}{219.91$^{***}$} & \multicolumn{1}{c}{45.73$^{***}$} & \multicolumn{1}{c}{5.98$^{***}$} \\
\hline
\hline \\[-1.8ex]
\multicolumn{7}{p{12.8cm}}{\textit{Note:} The regressions here estimate the effect of being convicted at trial on the probability of election for all candidates who have had their candidacy challenged under charges of electoral irregularities. Columns 1 and 4 display models not including individual candidate characteristics; columns 2 and 5 include age, gender, marital status, education level, political experience, and the amount spent in their campaign; columns 3 and 6 also include municipal, electoral, and party fixed-effects. I report robust standard errors for all specifications in this table. $^{*}$p$<$0.1; $^{**}$p$<$0.05; $^{***}$p$<$0.01} \\
\end{tabular}

\end{table}

\begin{table}[!htbp]
\centering
\caption[table6]{The Effect of Electoral Crime on the Total Vote Share}\label{tab:secondstageoutcome2}
\vspace{-10pt}

\centering
\scriptsize
\setlength{\tabcolsep}{-3pt}
\begin{tabular}{@{\extracolsep{4pt}}lD{.}{.}{-3} D{.}{.}{-3} D{.}{.}{-3} D{.}{.}{-3} D{.}{.}{-3} D{.}{.}{-3} }
\\[-1.8ex]\hline
\hline \\[-1.8ex]
 & \multicolumn{6}{c}{Outcome: Total Vote Share (in p.p.)} \\
\cline{2-7}
 \\[-1.8ex]
 & \multicolumn{1}{c}{OLS} & \multicolumn{1}{c}{OLS} & \multicolumn{1}{c}{OLS} & \multicolumn{1}{c}{IV} & \multicolumn{1}{c}{IV} & \multicolumn{1}{c}{IV} \\
\\[-1.8ex] & \multicolumn{1}{c}{(1)} & \multicolumn{1}{c}{(2)} & \multicolumn{1}{c}{(3)} & \multicolumn{1}{c}{(4)} & \multicolumn{1}{c}{(5)} & \multicolumn{1}{c}{(6)}\\
\hline \\[-1.8ex]
 Convicted on Appeal & -.097^{***} & -.055^{***} & -.075^{***} & -.153^{***} & -.101^{***} & -.129^{***} \\
  & (.003) & (.002) & (.003) & (.004) & (.003) & (.005) \\
  & & & & & & \\
\hline \\[-1.8ex]
Individual Controls & \multicolumn{1}{c}{-} & \multicolumn{1}{c}{Yes} & \multicolumn{1}{c}{Yes} & \multicolumn{1}{c}{-} & \multicolumn{1}{c}{Yes} & \multicolumn{1}{c}{Yes} \\
Fixed-Effects & \multicolumn{1}{c}{-} & \multicolumn{1}{c}{-} & \multicolumn{1}{c}{Yes} & \multicolumn{1}{c}{-} & \multicolumn{1}{c}{-} & \multicolumn{1}{c}{Yes} \\
\hline \\[-1.8ex]
Observations & \multicolumn{1}{c}{16,791} & \multicolumn{1}{c}{16,791} & \multicolumn{1}{c}{16,791} & \multicolumn{1}{c}{16,791} & \multicolumn{1}{c}{16,791} & \multicolumn{1}{c}{16,791} \\
\textit{F}-stat & \multicolumn{1}{c}{1635.53$^{***}$} & \multicolumn{1}{c}{618.08$^{***}$} & \multicolumn{1}{c}{7.91$^{***}$} & \multicolumn{1}{c}{1591.35$^{***}$} & \multicolumn{1}{c}{612.69$^{***}$} & \multicolumn{1}{c}{7.67$^{***}$} \\
\hline
\hline \\[-1.8ex]
\multicolumn{7}{p{13.5cm}}{\textit{Note:} The regressions here estimate the effect of being convicted at trial on the total vote share for all candidates who have had their candidacy challenged under charges of electoral irregularities. Columns 1 and 4 display models not including individual candidate characteristics; columns 2 and 5 include age, gender, marital status, education level, political experience, and the amount spent in their campaign; columns 3 and 6 also include municipal, electoral, and party fixed-effects. I report robust standard errors for all specifications in this table. $^{*}$p$<$0.1; $^{**}$p$<$0.05; $^{***}$p$<$0.01} \\
\end{tabular}

\end{table}

\begin{table}[!htbp]
\centering
\caption[table7]{The Effect of Electoral Crimes on the Vote Distance to Election Cutoff}\label{tab:secondstageoutcome3}
\vspace{-10pt}

\centering
\scriptsize
\setlength{\tabcolsep}{-2pt}
\begin{tabular}{@{\extracolsep{6pt}}lD{.}{.}{-3} D{.}{.}{-3} D{.}{.}{-3} D{.}{.}{-3} }
\\[-1.8ex]\hline
\hline \\[-1.8ex]
& \multicolumn{4}{c}{Outcome: Vote Distance to Election Cutoff (in p.p.)} \\
\cline{2-5}
\\[-1.8ex]
& \multicolumn{1}{c}{OLS} & \multicolumn{1}{c}{IV} & \multicolumn{1}{c}{OLS} & \multicolumn{1}{c}{IV} \\
\\[-1.8ex] & \multicolumn{1}{c}{(1)} & \multicolumn{1}{c}{(2)} & \multicolumn{1}{c}{(3)} & \multicolumn{1}{c}{(4)}\\
\hline \\[-1.8ex]
 Convicted at Trial & -.005^{***} & -.009^{***} & -.056^{***} & -.141^{***} \\
                    & (.000) & (.001) & (.013) & (.021) \\
                    & & & & \\
\hline \\[-1.8ex]
Individual Controls & \multicolumn{1}{c}{Yes} & \multicolumn{1}{c}{Yes} & \multicolumn{1}{c}{Yes} & \multicolumn{1}{c}{Yes} \\
Fixed-Effects       & \multicolumn{1}{c}{Yes} & \multicolumn{1}{c}{Yes} & \multicolumn{1}{c}{Yes} & \multicolumn{1}{c}{Yes} \\
Sample & \multicolumn{1}{c}{City Council} & \multicolumn{1}{c}{City Council} & \multicolumn{1}{c}{Mayor} & \multicolumn{1}{c}{Mayor} \\
\hline \\[-1.8ex]
Observations & \multicolumn{1}{c}{13,415} & \multicolumn{1}{c}{13,415} & \multicolumn{1}{c}{3,376} & \multicolumn{1}{c}{3,376} \\
\textit{F}-stat & \multicolumn{1}{c}{16.68$^{***}$} & \multicolumn{1}{c}{16.51$^{***}$} & \multicolumn{1}{c}{1.93$^{***}$} & \multicolumn{1}{c}{1.83$^{***}$} \\
\hline
\hline \\[-1.8ex]
\multicolumn{5}{p{10.25cm}}{\textit{Note:} The regressions here estimate the effect of being convicted at trial on the distance to the election cutoff for candidates who have had their candidacy challenged under charges of electoral irregularities. All models include individual candidate characteristics and municipal, electoral, and party fixed-effects. Since election rules differ by office type, I split the sample into city council candidates (columns 1 and 2) and mayor candidates (columns 3 and 4). I report robust standard errors for all specifications in this table. $^{*}$p$<$0.1; $^{**}$p$<$0.05; $^{***}$p$<$0.01} \\
\end{tabular}

\end{table}

\begin{table}[!htbp]
\centering
\caption[table8]{Bounded Estimates from \citet{OsterUnobservableSelectionCoefficient2019}}\label{tab:coefstability}
\vspace{-10pt}

\centering
\scriptsize
\setlength{\tabcolsep}{2pt}
\begin{tabular}{@{\extracolsep{6pt}}p{2cm}cc}
\\[-1.8ex]\hline
\hline \\[-1.8ex]
& (1) & (2) \\
\\[-1.8ex]
% \cline{2-3} \\[-1.8ex]
& $R_\text{max} = 1.3\cdot\tilde{R}$ & $R_\text{max} = 1$ \\
\\ [-1.8ex]
\hline
\\ [-1.8ex]
$\beta_\text{trial}$& [$-$0.063,$-$0.072] & [$-$0.063,$-$0.081] \\
\\[-1.8ex]
$\beta_\text{appeals}$& [$-$0.053,$-$0.065] & [$-$0.053,$-$0.075] \\
\\[-1.8ex]
\hline
\hline
\\[-1.8ex]
\multicolumn{3}{p{6.7cm}}{\textit{Note:} This table depicts the bounded estimates for [$\tilde{\beta},\beta^{*}$] in the two sets of regressions. $\tilde{\beta}$ is the coefficient for each conviction variable in the multivariate model with individual controls and party, municipal, and election fixed-effects. $\beta^{*}$ is the bias-adjusted coefficient for each conviction variable in the same multivariate model using $\delta = 1$ as the degree of selection on unobservables and under $R_\text{max}$ values on the top row.}
\end{tabular}

\end{table}

\begin{table}[!htbp]
\centering
\caption[table9]{Heterogeneous Sentencing across Trial and Appeals}\label{tab:heterogeneoussentencing}
\vspace{-10pt}

\centering
\scriptsize
\setlength{\tabcolsep}{0pt}
\begin{tabular}{@{\extracolsep{3pt}}lD{.}{.}{-3} D{.}{.}{-3} D{.}{.}{-3} D{.}{.}{-3} D{.}{.}{-3} D{.}{.}{-3}}
\\[-1.8ex]\hline
\hline \\[-1.8ex]
& \multicolumn{1}{c}{(1)} & \multicolumn{1}{c}{(2)} & \multicolumn{1}{c}{(3)} & \multicolumn{1}{c}{(4)} & \multicolumn{1}{c}{(5)} & \multicolumn{1}{c}{(6)} \T \B \\

& \multicolumn{1}{c}{$\beta$\textsubscript{trial}} & \multicolumn{1}{c}{$\beta$\textsubscript{appeals}} & \multicolumn{1}{c}{$\beta$\textsubscript{difference}} & \multicolumn{1}{c}{s.e.} & \multicolumn{1}{c}{\textit{t}-stat} & \multicolumn{1}{c}{\textit{p}-value} \T \B \\
\hline \\[-1.8ex]
Elected to Office                           & -.148 & -.124 & -.024 & .053 & -.456  & .649 \\
Age                                         & .003  & .001  & .001  & .002 & .731   & .465 \\
Male                                        & .004  & .003  & .001  & .013 & .040   & .968 \\
Political Experience                        & -.036 & -.068 & .032  & .090 & .361   & .718 \\
Campaign Expenditures (in R\$)              & -.044 & -.033 & -.011 & .008 & -1.335 & .182 \\
Marital Status:                             & & & & & & \\
\hspace{10pt}Divorced                       & .057  & .039  & .017  & .035 & .495   & .621 \\
\hspace{10pt}Legally Divorced               & .025  & .044  & -.019 & .067 & -.286  & .775 \\
\hspace{10pt}Single                         & .074  & .050  & .023  & .032 & .733   & .464 \\
\hspace{10pt}Widowed                        & -.014 & -.024 & .009  & .060 & .159   & .873 \\
Educational Levels:                         & & & & & & \\
\hspace{10pt}Completed ES/MS                & -.178 & -.304 & .127  & .104 & 1.216  & .224 \\
\hspace{10pt}Incomplete ES/MS               & -.208 & -.308 & .100  & .158 & .633   & .527 \\
\hspace{10pt}Can Read and Write             & -.241 & -.326 & .086  & .234 & .366   & .714 \\
\hspace{10pt}Completed HS                   & -.185 & -.317 & .133  & .088 & 1.504  & .133 \\
\hspace{10pt}Incomplete HS                  & -.234 & -.312 & .078  & .158 & .495   & .620 \\
\hspace{10pt}Completed College              & -.227 & -.363 & .136  & .096 & 1.411  & .158 \\
\hspace{10pt}Incomplete College             & -.201 & -.319 & .118  & .119 & .994   & .320 \\

\hline
\hline \\[-1.8ex]
\multicolumn{7}{p{11.3cm}}{\textit{Note:} In this table, I report the coefficients of two regressions using the same covariates on the probability of receiving an unfavorable ruling at trial (column 1) and on appeals (column 2). I then recover the distributions of the differences in betas and test H0: $\beta$\textsubscript{difference} = 0 for all covariates in the regressions (columns 3-6). Robust standard errors are clustered at the municipal-election pair level (equivalent to the judge-level error shared by all candidates in one municipality during one election period); party-fixed effects are included in both regressions but are not reported here.} \\
\end{tabular}

\end{table}

\begin{table}[!htbp]
\centering
\caption[table10]{The Effect of Electoral Crimes on Voter Engagement}\label{tab:voterbehavior}
\vspace{-10pt}

\centering
\scriptsize
\setlength{\tabcolsep}{-2pt}
\begin{tabular}{@{\extracolsep{6pt}}lcccc}
\\[-1.8ex]\hline
\hline \\[-1.8ex]
& \multicolumn{2}{c}{Party-Level} & \multicolumn{2}{c}{Election-Level} \\
\\[-1.8ex]
\cline{2-3} \cline{4-5}
\\[-1.8ex] & \multicolumn{1}{p{1.2cm}}{\centering Outcome: Voter Turnout (percent)} & \multicolumn{1}{p{1.2cm}}{\centering Outcome: Invalid Votes (percent)} & \multicolumn{1}{p{1.2cm}}{\centering Outcome: Voter Turnout (percent)} & \multicolumn{1}{p{1.2cm}}{\centering Outcome: Invalid Votes (percent)} \\
\\[-1.8ex] & (1) & (2) & (3) & (4) \\
\hline \\[-1.8ex]
Share of Candidacies          & .015 & \hspace{12pt}.276$^{***}$ & \hspace{-9pt} $-$.004 & \hspace{3pt}.183$^{*}$ \\
\hspace{4pt} Invalid at Trial & (.009) & (.085) & (.006) & (.083) \\
                              & & & & \\
\hline \\[-1.8ex]
Individual Controls & - & - & - & - \\
Fixed-Effects       & Yes & Yes & Yes & Yes \\
\hline \\[-1.8ex]
Observations        & 8,950 & 8,950 & 5,496 & 5,496 \\
\textit{F}-stat & 434.2$^{***}$ & 1298.9$^{***}$ & 145.7$^{***}$ & 354.3$^{***}$ \\
\hline
\hline \\[-1.8ex]
\multicolumn{5}{p{8cm}}{\textit{Note:} The regressions here estimate the effect of the share of candidates convicted at trial overall the total office vacancies on voter turnout and the number of invalid votes (both logged). I aggregate observations up to the party and election level and control for municipality and election year fixed-effects. I report robust standard errors, clustered by elections and municipalities, for all specifications in this table. $^{*}$p$<$0.1; $^{**}$p$<$0.05; $^{***}$p$<$0.01} \\
\end{tabular}

\end{table}

\begin{table}[!htbp]
\centering
\caption[table11]{Campaign Expenditure Across Ruling Group}\label{tab:candidatebehavior}
\vspace{-10pt}

\centering
\scriptsize
\setlength{\tabcolsep}{0pt}
\begin{tabular}{@{\extracolsep{6pt}}lcccc}
\\[-1.8ex]\hline
\hline \\[-1.8ex]
& \multicolumn{2}{p{3cm}}{\centering \textit{Mean Campaign Spending in Ruling Group (in R\$)}} & & \\
\cline{2-3} \\[-1.8ex]
\textit{Stage} & Favorable & Unfavorable & \textit{t}-stat & \textit{p}-value \T \B \\
\hline \\[-1.8ex]
\hspace{1pt} Trial        & 79,099 & 35,402 & 5.79 & .000 \T \B \\
& \tiny [7,095] & \tiny [9,696] & & \\
\hspace{1pt} Appeals      & 71,522 & 36,001 & 4.83 & .000 \T \B \\
& \tiny [8,445] & \tiny [8,346] & & \\
& & & & \\
\textit{Unfavorable Ruling} & Affirmed & Reversed & \textit{t}-stat & \textit{p}-value \T \B \\
\hline \\[-1.8ex]
\hspace{1pt} Trial        & 39,004 & 34,324 & 0.63 & 0.53 \T \B \\
 & \tiny [7,461] & \tiny [2,235] & & \\
\\[-1.8ex]\hline
\hline \\ [-1.8ex]
\multicolumn{5}{p{7.9cm}}{\textit{Note:} This table reports t-tests across different subsamples of candidates. The number of observations in each group is reported inside the squared brackets.}
\end{tabular}

\end{table}

% \begin{figure}[!htbp]
% \centering
% \caption[figure4]{Simulation of IV Point Estimates}\label{fig:weakinstruments}
% \vspace{-10pt}
% \hspace{-30pt}\includegraphics[scale = .6]{plots/weakinstruments.pdf}
% \end{figure}

\begin{table}[!htbp]
\centering
\caption[table12]{Voter Sophistication and Benefit of Rule-Breaking}\label{tab:htehypotheses}
\vspace{-10pt}

\centering
\scriptsize
\setlength{\tabcolsep}{2pt}
\begin{tabular}{@{\extracolsep{6pt}}lccc}
\\[-1.8ex]\hline
\hline \\[-1.8ex]
& & \multicolumn{2}{c}{$\rho_{1}$: Severe Violation} \\
\\[-.9ex]
% \cline{2-3} \\[-1.8ex]
& & \multicolumn{1}{c|}{$\rho_{1} = 0$} & $\rho_{1} > 0$ \\ [.9ex]
\multirow{6}{*}{\parbox{3.5cm}{$\rho_{2}$: Convicted at Trial $\times$ \\ \raggedleft Severe Violation}} & \multirow{3}{*}{$\rho_{2} = 0$} & \multicolumn{1}{m{6cm}|}{\raggedright \textbf{1.} Violation carries no electoral benefit.} & \multicolumn{1}{m{6cm}}{\raggedright \textbf{1.} Violation helps candidate get elected.} \\
& & \multicolumn{1}{m{6cm}|}{\raggedright \textbf{2.} Voters impose same penalty for different electoral violations.} & \multicolumn{1}{m{6cm}}{\raggedright \textbf{2.} Voters impose same penalty for different electoral violations.} \\
& & \multicolumn{1}{m{6cm}|}{} & \\
\cline{2-4}
& & \multicolumn{1}{m{6cm}|}{} & \\
& \multirow{3}{*}{$\rho_{2} < 0$} & \multicolumn{1}{m{6cm}|}{\raggedright \textbf{1.} Violation carries no electoral benefit.} & \multicolumn{1}{m{6cm}}{\raggedright \textbf{1.} Violation helps candidate get elected.} \\
& & \multicolumn{1}{m{6cm}|}{\raggedright \textbf{2.} Voters impose harsher electoral penalties for severe violations.} & \multicolumn{1}{m{6cm}}{\raggedright \textbf{2.} Voters impose harsher electoral penalties for severe violations.} \\ [3ex]
\\ [-1.8ex]
\hline
\hline \\[-1.8ex]
\end{tabular}

\end{table}

\begin{table}[!htbp]
\centering
\caption[table13]{Heterogeneous Effect of Electoral Ruling}\label{tab:hte}
\vspace{-10pt}

\centering
\scriptsize
\setlength{\tabcolsep}{-2pt}
\begin{tabular}{@{\extracolsep{6pt}}lcccc}
\\[-1.8ex]\hline
\hline \\[-1.8ex]
& \multicolumn{2}{p{3.6cm}}{\centering Full Sample} & \multicolumn{1}{p{1.8cm}}{\centering City Councilor} & \multicolumn{1}{p{1.8cm}}{\centering Mayor} \\
\\[-1.8ex]
\cline{2-3} \cline{4-4} \cline{5-5} \\[-1.8ex]
& \multicolumn{1}{p{1.8cm}}{\centering Outcome: Probability of Election} & \multicolumn{1}{p{1.8cm}}{\centering Outcome: Vote Share \quad (in p.p.)} & \multicolumn{1}{p{1.8cm}}{\centering Outcome: Vote Distance to Cutoff \quad (in p.p.)} & \multicolumn{1}{p{1.8cm}}{\centering Outcome: Vote Distance to Cutoff \quad (in p.p.)} \\
\\[-1.8ex] & (1) & (2) & (3) & (4) \\
\hline \\[-1.8ex]
Convicted at Trial                         & \hspace{5pt}$-$.062$^{***}$ & \hspace{5pt}$-$.095$^{***}$ & \hspace{4pt}$-$.009$^{***}$ & \hspace{5pt}$-$.189$^{***}$ \\
                                           & (.016) & (.006) & (.001) & (.023) \\
                                           & & & & \\
Severe Violation                           & \hspace{11pt}.069$^{***}$ & \hspace{11pt}.052$^{***}$ & .000 & \hspace{5pt}$-$.040$^{***}$ \\
                                           & (.019) & (.008) & (.001) & (.015) \\
                                           & & & & \\
Convicted at Trial                         & \hspace{2pt}$-$.065$^{**}$ & \hspace{5pt}$-$.058$^{***}$ & .001 & \hspace{11pt}.084$^{***}$ \\
\hspace{4pt} $\times$ Severe Violation     & (.027) & (.011) & (.001) & (.029) \\
                                           & & & & \\
\hline \\[-1.8ex]
Individual Controls & \multicolumn{1}{c}{Yes} & \multicolumn{1}{c}{Yes} & \multicolumn{1}{c}{Yes} & \multicolumn{1}{c}{Yes} \\
Fixed-Effects       & \multicolumn{1}{c}{Yes} & \multicolumn{1}{c}{Yes} & \multicolumn{1}{c}{Yes} & \multicolumn{1}{c}{Yes} \\
\hline \\[-1.8ex]
Observations & \multicolumn{1}{c}{11,095} & \multicolumn{1}{c}{11,095} & \multicolumn{1}{c}{9,070} & \multicolumn{1}{c}{2,025} \\
\textit{F}-stat & \multicolumn{1}{c}{5.92$^{***}$} & \multicolumn{1}{c}{8.12$^{***}$} & \multicolumn{1}{c}{23.50$^{***}$} & \multicolumn{1}{c}{1.62$^{***}$} \\
\hline
\hline \\[-1.8ex]
\multicolumn{5}{p{10cm}}{\textit{Note:} The regressions here include the severity of the accusation brought against candidates running for municipal office. I recover the accusations from court documents and identify ruling type using linear support-vector machine classification (details in appendix
A). In columns 1-4, I report the coefficients on ruling outcome (row 1), type (row 2), and their interaction (row 3). All regressions include municipal, electoral, and party fixed-effects. Robust standard errors are displayed inside parentheses. $^{*}$p$<$0.1; $^{**}$p$<$0.05; $^{***}$p$<$0.01} \\
\end{tabular}

\end{table}

\pagebreak

\appendix
\section{Appendix: Electoral Ruling Classification} \label{sec:machinelearning_paper1}

The Brazilian Electoral Court (TSE) is responsible for authorizing individual candidacies for elected office. Every cycle, all candidates submit proper documentation to the court, ahead of elections, and the court authorizes or dismisses candidacies based on statutory electoral law. I discuss this mechanism at length in section \ref{sec:background_paper1}. These decisions are actual judicial sentences issued by judges at trial and appellate (electoral) courts. These sentences are public documents, and are available both on paper at electoral district courtrooms across the country or as electronic documents on the TSE's website. I use these data in this project. Due to TSE's online database management limitations, however, I can only recover sentences for some candidates in my sample. There is no reason to believe there is any selection bias in the disclosure of these documents: almost no sentences are available for the 2004 and 2008 elections, but all sentences are available for the 2012 and 2016 elections, indicating an information technology, rather than a political favoritism, issue.

To conduct the heterogeneous effect analysis in section \ref{sec:hte_paper1}, I classify these sentences into severe or trivial rule-breaking according to electoral law in Brazil. There are eight reasons preventing candidates from running for office:
\begin{enumerate*}[label = \textbf{(\arabic*)}]
  \item \underline{individual documentation is incomplete}: candidates have not included their social security numbers, identification card numbers, photos, or other information in candidacy applications;
  \item \underline{party documentation is incomplete}: parties might have not presented financial records, or coalition did not meet legal requirements;
  \item \underline{candidacy impeachment}: the electoral prosecutor or private parties (opponents, political parties) filed, and were granted, a request for impeachment based on violations to electoral law;
  \item \underline{use of public office for electoral gain}: public officials have used their office for direct benefit (when they are on the ticket) or indirect benefit (when they do it to support someone else, usually a political ally);
  \item \underline{illegal campaign spending}: candidates have used funds for activities forbidden by electoral law or have spent beyond spending limits;
  \item \underline{vote buying}: candidates have paid, in cash or in kind, individuals in exchange for their votes;
  \item \underline{abuse of economic power}: candidates channelled campaign spending via private parties, usually business executives who support candidates and campaign on their behalf;
  \item \underline{previous criminal convictions}: candidates had been convicted at trial, and their conviction was affirmed on appeals, for crimes in the past (corruption, murder, abuse of power, and others).
\end{enumerate*} Reasons 1 and 2 are classified as trivial rule-breaking, since they are related to procedural reasons for dismissal of candidacies. Individuals or parties could have easily fixed those by submitting or keeping proper documentation. All other reasons are instances of severe electoral violations; candidates actively engaged in illegal actions either before or when running for public office. All of these regulations are codified in electoral legislation summarized throughout the paper.

Importantly, TSE created these categories and has been publishing individual-level information for rejections since 2014. I use these categories as classes, and words in each sentence as features, for training machine classification algorithms. First, I group categories 1 and 2 as trivial rule-breaking, and categories 3-8 as severe rule-breaking. The imbalance in the classes justifies such approach: some categories make up less than one percent of the sample, thus the algorithm would not meaningfully predict them on such a small set of observations. Second, classes 3-8 are based on the Clean Records Act of 2010 and the Law of Elections of 1997; their provisions and punishment are relatively standard, making individual violations less relevant for criminal decisions. Individuals are usually indicted on multiple counts of breach of both statutes, therefore the relevant break in behavior occurs between trivial and non-trivial actions. After this process, the ratio of classes is 80/20 for procedural and severe electoral violations.

\begin{figure}[htbp]
\centering
\caption[figure5]{Accuracy Scores}\label{fig:accuracy}
\vspace{-10pt}
\hspace{-30pt}\includegraphics[scale = .7]{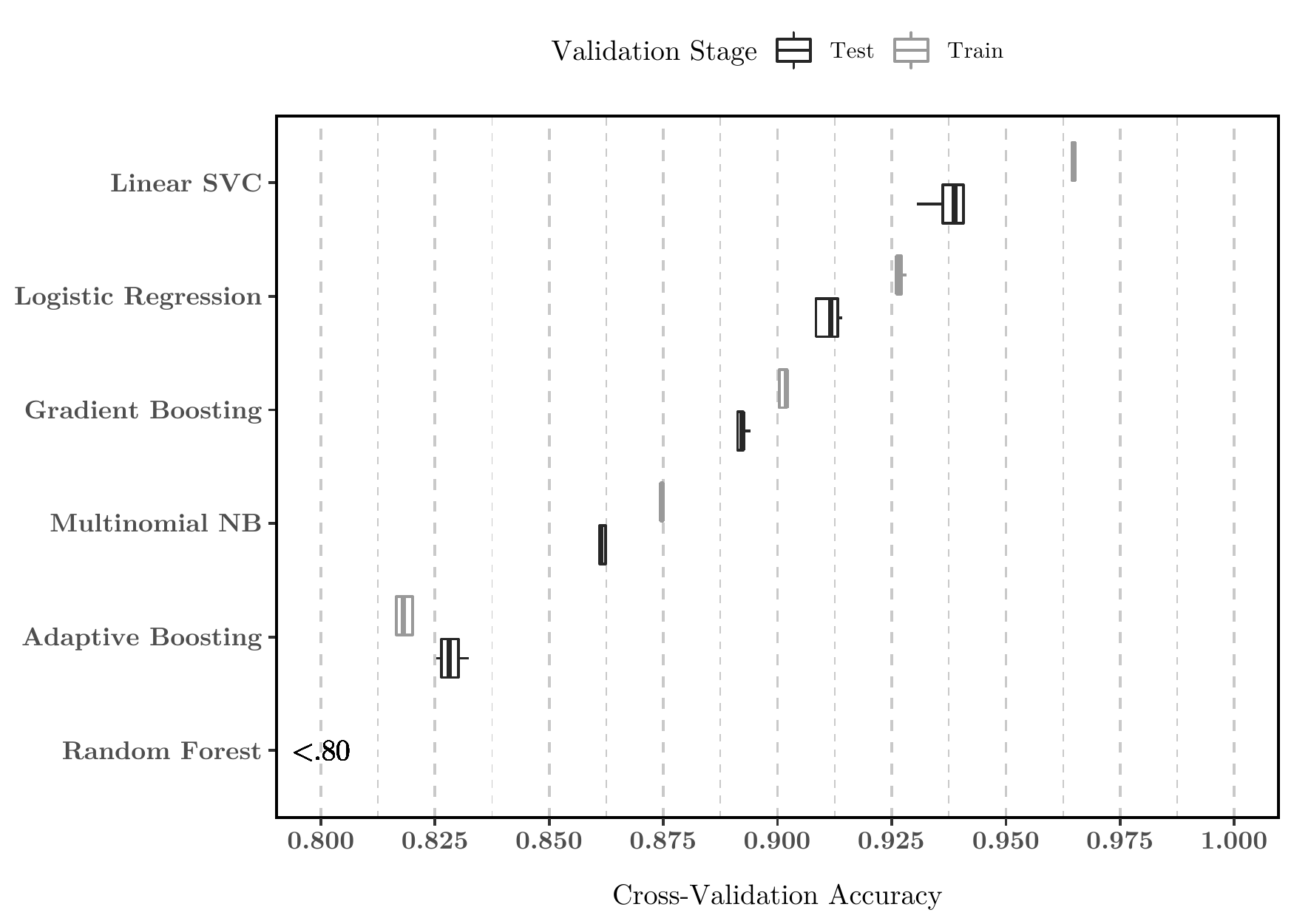}
\end{figure}

The features used for classification come from each judicial sentence available on the TSE's website, which is almost exclusively composed of sentences in 2012 and 2016 (over 99 percent of all sample). I process the text in each sentence by eliminating stopwords and computing the term frequency-inverse document frequency (TF-IDF) for every unigram and bigram appearing more than five times in all sentences. The final dataset contains more than 278,992 features (variables) and 57,478 sentences (rows).  Candidates can have multiple sentences if they appealed their cases to higher courts. Using these data, I train six algorithms on 80 percent of 2016 data and test on a 20 percent hold-out to predict 2012 classes. The classification algorithms are: Multinomial Naive Bayes Classifier (NB), Logistic Regression, Linear Support-Vector Classification (Linear SVC), Random Forest, Adaptive Boosting, and Gradient Boosting. I compute their accuracy scores and area under the curve (AUC), and implement five-fold cross validation at training. Figure \ref{fig:accuracy} displays mean accuracy across validation folds. I also report the mean area under the curve across validation folds in panel A of table \ref{tab:auc}. I also run a Deep Neural Networks model that uses 300-dimension word vectors for each of the 16,396 unique words, but I omit such model here because it performs much than the others (85 percent accuracy). The best classification algorithm is the Linear SVC model, with an average of 93 percent accuracy across training folds and .98 area under the curve (AUC). In panel B of table \ref{tab:auc}, I report the hold-out classification results, and the best performing model is also Linear SVC, with 88 percent accuracy and .731 area under the curve. Therefore, I use such model to predict the reasons for convicting candidates in 2012. These predictions then become the heterogeneous treatment variable in section \ref{sec:hte_paper1}, which break down the conviction effect into two types: conviction for severe or trivial electoral crimes.

\clearpage

\begin{table}[htbp]
\centering
\caption[table14]{Validation Accuracy and AUC Scores}\label{tab:auc}
\vspace{-10pt}

\centering
\scriptsize
\setlength{\tabcolsep}{0pt}
\begin{tabular}{@{\extracolsep{10pt}}lp{1.5cm}p{1.5cm}l@{\extracolsep{-2pt}}p{1.5cm}p{1.5cm}}
\\[-1.8ex]\hline
\hline \\[-1.8ex]
\multicolumn{3}{c}{\emph{Panel A: 5-fold Cross-Validation Sample}} & \multicolumn{3}{c}{\emph{Panel B: Hold-Out Sample}} \\
\cline{1-3} \cline{4-6}\\
\emph{Model} & \multicolumn{1}{c}{Accuracy} & \multicolumn{1}{c}{AUC} & \emph{Model} & \multicolumn{1}{c}{Accuracy} & \multicolumn{1}{c}{AUC} \\
& \multicolumn{1}{c}{(mean)} & \multicolumn{1}{c}{(mean)} & & & \\
\hline \\[-1.8ex]
\\[-1.8ex]
\hspace{2pt} 1. Linear SVC          & \multicolumn{1}{c}{0.930} & \multicolumn{1}{c}{0.980} & \hspace{2pt} 1. Linear SVC          & \multicolumn{1}{c}{0.880} & \multicolumn{1}{c}{0.731} \\
\hspace{2pt} 2. Logistic Regression & \multicolumn{1}{c}{0.892} & \multicolumn{1}{c}{0.953} & \hspace{2pt} 2. Random Forest       & \multicolumn{1}{c}{0.879} & \multicolumn{1}{c}{0.714} \\
\hspace{2pt} 3. Multinomial NB      & \multicolumn{1}{c}{0.852} & \multicolumn{1}{c}{0.925} & \hspace{2pt} 3. Logistic Regression & \multicolumn{1}{c}{0.859} & \multicolumn{1}{c}{0.706} \\
\hspace{2pt} 4. Gradient Boosting   & \multicolumn{1}{c}{0.802} & \multicolumn{1}{c}{0.894} & \hspace{2pt} 4. Multinomial NB      & \multicolumn{1}{c}{0.842} & \multicolumn{1}{c}{0.690} \\
\hspace{2pt} 5. Adaptive Boosting   & \multicolumn{1}{c}{0.747} & \multicolumn{1}{c}{0.849} & \hspace{2pt} 5. Gradient Boosting   & \multicolumn{1}{c}{0.841} & \multicolumn{1}{c}{0.666} \\
\hspace{2pt} 6. Random Forest       & \multicolumn{1}{c}{0.617} & \multicolumn{1}{c}{0.789} & \hspace{2pt} 6. Adaptive Boosting   & \multicolumn{1}{c}{0.788} & \multicolumn{1}{c}{0.628} \\
\\[-1.8ex]
\hline \\[-1.8ex]
\hline
\end{tabular}

\end{table}

\end{document}